\documentclass{aa} %
\usepackage{amsmath} %
\usepackage{amssymb} 
\usepackage{graphics} %
\usepackage{subfigure}
\usepackage{color} %
\usepackage{float}  
\usepackage{natbib} %
\usepackage{graphicx} 
\usepackage{etex}
\usepackage{txfonts}

\newcommand{\lya}{Ly$\alpha$}
\newcommand{\flya}{$F_{\rm Ly\alpha}$}
\newcommand{\llya}{$L_{\rm Ly\alpha}$}

\newcommand{\ew}{EW$_{\rm 0}$}

\newcommand{\muv}{$M_{\rm UV}$}

\newcommand{\udft}{\textsf{udf-10}}
\newcommand{\mosaic}{\textsf{mosaic}}
\newcommand{\unil}{erg s$^{-1}$}

\newcommand{\llyaunil}{(\llya/erg s$^{-1}$)}

\begin{document}

   \title{The MUSE Hubble Ultra Deep Field Survey}

   \subtitle{X. \lya\ Equivalent Widths at $2.9 < z < 6.6$}

   \author{
        T. Hashimoto\inst{1,2,3,4},
        T. Garel\inst{1}, 
        B. Guiderdoni\inst{1},
        A. B. Drake\inst{1},
        R. Bacon\inst{1},
        J. Blaizot\inst{1}, 
        J. Richard\inst{1},
        F. Leclercq\inst{1},    
        H. Inami\inst{1}, 
        A. Verhamme\inst{1, 5}, 
        R. Bouwens\inst{6}, 
        J. Brinchmann\inst{6, 7}, 
        S. Cantalupo\inst{8}, 
        M. Carollo\inst{8}
        J. Caruana\inst{9, 10}, 
        E. C. Herenz\inst{11}, 
        J. Kerutt\inst{12}, 
        R. A. Marino\inst{8}, 
        P. Mitchell\inst{1}, 
        and 
        J. Schaye\inst{6}
          }

\offprints{thashimoto@est.osaka-sandai.ac.jp}
\institute{
Univ Lyon, Univ Lyon1, Ens de Lyon, CNRS, Centre de Recherche Astrophysique de Lyon \\
UMR5574, F-69230, Saint-Genis-Laval, France 
\and
Department of Astronomy, Graduate School of Science, 
The University of Tokyo, Tokyo 113-0033, Japan 
\and 
National Astronomical Observatory of Japan, 2-21-1 Osawa, Mitaka, Tokyo 181-8588, Japan
\and 
College of General Education, Osaka Sangyo University, 
3-1-1 Nakagaito, Daito, Osaka 574-8530, Japan
\and
Observatoire de Gen\`eve, Universit\'e de Gen\`eve, 51 Ch. des Maillettes, 1290 Versoix, Switzerland
\and 
Leiden Observatory, P.O. Box 9513, NL-2300 RA Leiden, The Netherlands
\and 
Instituto de Astrof{\'\i}sica e Ci{\^e}ncias do Espa\c{c}o, Universidade do Porto, CAUP, Rua das Estrelas, PT4150-762 Porto, Portugal
\and 
Institute for Astronomy, Department of Physics, ETH Z\"urich, Wolfgang-Pauli-Strasse 27, 8093, Z\"urich, Switzerland
\and 
Department of Physics, University of Malta, Msida MSD 2080, Malta
\and 
Institute for Space Sciences \& Astronomy, University of Malta, Msida MSD 2080, Malta
\and 
Department of Astronomy, Stockholm University, AlbaNova University Centre, SE-106 91, Stockholm, Sweden
\and
Leibniz-Institut f\"{u}r Astrophysik Potsdam (AIP), An der Sternwarte 16, 14482, Potsdam, Germany 
}
\date{Received date / Accepted date}
\authorrunning{Hashimoto et al.}{ }
\titlerunning{\lya\ equivalent widths for \lya\ emitters at $2.9 < z < 6.6$}{ }

   \date{Received July, 2017; accepted Oct, 2017}

\abstract
{
We present rest-frame \lya\ equivalent widths (\ew) of $417$ \lya\ emitters (LAEs) 
detected with Multi Unit Spectroscopic Explorer (MUSE) on the Very Large Telescope (VLT) at $2.9 < z < 6.6$ in the Hubble Ultra Deep Field. 
Based on the deep MUSE spectroscopy and ancillary Hubble Space Telescope (HST) photometry data, 
we carefully measured \ew\ values taking into account extended \lya\ emission 
and UV continuum slopes ($\beta$). 
Our LAEs reach unprecedented depths, both in \lya\ luminosities and UV absolute magnitudes, 
from log \llyaunil \ $\sim41.0$ to $43.0$ 
and from \muv\ $\sim -16$ to $-21$ ($0.01-1.0$ $L^{*}_{\rm z=3}$). 
The \ew\ values span the range of $\sim 5$ to $240$ \AA\ or larger, 
and their distribution can be well fitted by an exponential law 
$N = N_{\rm 0}$ exp($-$\ew/$w_{\rm 0}$). 
Owing to the high dynamic range in \muv, 
we find that the scale factor, $w_{\rm 0}$, depends on \muv\ 
in the sense that including fainter \muv\ objects increases $w_{\rm 0}$, i.e., 
the Ando effect. 
The results indicate that selection functions affect the \ew\ scale factor. 
Taking these effects into account, 
we find that our $w_{\rm 0}$ values are consistent with those in the literature 
within $1\sigma$ uncertainties at $2.9 < z < 6.6$ at a given threshold of \muv\ and \llya.
Interestingly, we find $12$ objects 
with \ew\ $>200$ \AA\ above $1\sigma$ uncertainties. 
Two of these $12$ LAEs show signatures of merger or AGN activity: 
the weak C{\sc iv} $\lambda 1549$ emission line. 
For the remaining $10$ very large \ew\ LAEs, 
we find that the \ew\ values can be reproduced by 
young stellar ages ($< 100$ Myr)  and low metallicities ($\lesssim 0.02$ $Z_{\rm \odot}$). 
Otherwise, at least part of the \lya\ emission in these LAEs needs to arise from 
anisotropic radiative transfer effects, fluorescence by hidden AGN or quasi-stellar object activity, 
or gravitational cooling.
}
\keywords{
cosmology: observations --- 
galaxies: evolution ---
galaxies: formation --- 
galaxies: high-redshift
 }
\maketitle
%

\section{Introduction}
\label{sec:introduction}

\lya\ emitters (LAEs) are galaxies selected by virtue of their strong \lya\ emission. 
Numerous LAEs have been discovered using 
the narrowband technique
(e.g., \citealt{cowie1998, rhoads2000, shimasaku2006, gronwall2007, ouchi2008, ouchi2010, cowie2011, shibuya2017}) 
or direct spectroscopic searches 
(e.g., \citealt{shapley2003, santos2004, rauch2008, cassata2015}). 

Apart from redshift determinations of high $z$ galaxies 
(\citealt{finkelstein2013, schenker2014, zitrin2015}), 
the \lya\ line is useful to examine stellar populations of galaxies 
(e.g., \citealt{schaerer2003, dijkstra2014}) 
and can be used to probe the distribution and kinematics of cool gas in and around galaxies 
(e.g., \citealt{mas-hesse2003, verhamme2006, steidel2011}). 
However, interpretations are often complicated 
because of the intricate radiative transfer of the \lya\ line 
(theoretical studies: e.g., \citealt{dijkstra2006, laursen2011,verhamme2006, verhamme2012, gronke2016b}; observational studies: e.g., \citealt{hayes2013, hayes2014, hashimoto2015, herenz2016}). 

A widely used tracer of these processes is the rest-frame \lya\ equivalent width (\ew). 
Based on stellar synthesis models, 
\cite{schaerer2003} and \cite{raiter2010} showed that 
\ew\ becomes intrinsically larger 
for galaxies with young stellar ages, low metallicities, 
or a top-heavy initial mass function (IMF). 
According to these theoretical studies, 
it is possible to reproduce values of \ew\ $\lesssim200$ \AA\ 
with models of stellar populations with a normal Salpeter IMF (\citealt{salpeter1955}) 
and solar metallicity ($1.0$ $Z_{\rm \odot}$) 
(cf. \citealt{charlot1993, malhotra2002}).

According to previous narrowband surveys, 
a significant fraction of LAEs ($10-40$ \%) seem to show very large \ew\ $\gtrsim 200$\AA\ 
(e.g., \citealt{malhotra2002, shimasaku2006, ouchi2008}). 
Very large \ew\ LAEs are also spectroscopically identified in some studies 
(e.g., \citealt{dawson2004, adams2011, kashikawa2012, hashimoto2017}). 
According to stellar synthesis models of \cite{schaerer2003} and \cite{raiter2010}, 
the very large \ew\ values can be reproduced by either a top-heavy IMF, very young stars ($\lesssim$ 10 Myr) , 
or very low metallicity stars ($\lesssim$ 0.02 $Z_{\rm \odot}$). 
Thus, very large \ew\ LAEs are important as candidates 
of galaxies hosting metal-free stars (Population III stars; hereafter PopIII stars). 
Alternatively, the very large \ew\ values can be reproduced by either \lya\ fluorescence due to a hard-ultraviolet spectrum produced by in situ AGN activity or nearby quasi-stellar objects (QSOs;
e.g., \citealt{malhotra2002, cantalupo2012}) 
or  cooling radiation from shock-heated gas (e.g., \citealt{rosdahl2012, yajima2012}).

However, there are three problems with estimates of \ew\ from previous studies. 
First, it is now known that \lya\ emission is significantly extended compared with UV emission 
(e.g., \citealt{steidel2011, hayes2013, momose2014, wisotzki2016, patricio2016, sobral2017, Leclercq2017}). 
Thus, previous studies had difficulty in estimating total \lya\ fluxes. 
For spectroscopic studies, as \cite{rauch2008} pointed out, 
the slit losses can be up to $20-50\%$ of the total fluxes. 
Second, because LAEs have faint continua, 
the continuum fluxes are difficult to measure from spectroscopic data.  
Thus, most studies have estimated 
continuum fluxes at $1216$\AA\ from broadband photometry in the wavelength range 
redward of the Ly$\alpha$ line.  
In this calculation, 
a flat UV continuum slope, $\beta=-2.0$, is typically assumed, 
where $\beta$ is defined as f$_{\lambda}$ = $\lambda^{\beta}$ 
(e.g., \citealt{malhotra2002, shimasaku2006, guaita2011}), 
although several studies have simultaneously 
derived $\beta$ and \ew\ (e.g., \citealt{blanc2011, jiang2013, hashimoto2017}). 
Therefore, most previous studies suffer from systematic uncertainties in the continuum fluxes at $1216$\AA\ 
and in \ew.
Finally, a proper association of \lya\ emission to UV counterparts is sometimes difficult 
because of the source crowding in the projected sky. 
This is particularly the case for ground-based telescopes 
where the point spread function (PSF) is too large to deblend crowded sources 
(see also \citealt{Brinchmann2017}). 
Wrong associations can cause inaccurate measurements of \ew. 
These problems mean that both the narrowband technique and slit spectroscopy 
suffer from their own shortcomings. 

To address these problems, 
we present a new sample of LAEs obtained from deep observations 
with the Multi Unit Spectroscopic Explorer (MUSE; \citealt{bacon2010}) on the Very Large Telescope (VLT) 
in the Hubble Ultra Deep Field (UDF; \citealt{beckwith2006}) . 
The UDF is equipped with extremely deep photometric data, 
which are useful to constrain accurate continuum fluxes at $1216$ \AA. 
The capabilities of this unique integral field unit (IFU) spectrograph, 
in particular its high sensitivity and spectral/spatial resolution, together with the HST data 
enable us to produce a homogeneous sample of faint LAEs with unprecedented depth. 

In this study, we focus on two controversial issues: first, the evolution of the \ew\ distribution between $z=2.9$ and $6.6,$ 
and second, the existence of very large \ew\ LAEs. 

Regarding the first point, 
many observational studies have examined the \ew\ distribution,
and several of these studies have also investigated the evolution of the distribution. 
The distribution is often expressed 
as an exponential law 
$N$ = $N_{\rm 0}$ exp($-$\ew/$w_{\rm 0}$), 
where $w_{\rm 0}$ is the scale factor of \ew\ 
(e.g., \citealt{gronwall2007, nilsson2009, guaita2010, ciardullo2012, zheng2014, oyarzun2016, oyarzun2017, shibuya2017}). 
Based on a compiled sample of LAEs at $0 < z < 6$ from the literature, 
\cite{zhenya_zheng2014} claimed that 
$w_{\rm 0}$ becomes large at high $z$ 
(see also \citealt{ciardullo2012} who found similar redshift evolution at $2 < z < 3$). 
These results suggest that large \ew\ LAEs are more common at higher $z$, 
which may be consistent with the evolution 
of the fraction of strong \lya\ emission among 
dropout galaxies (e.g., \citealt{stark2010, cassata2015}). 
However, the results on the redshift evolution  are based on a compiled sample 
that comprises LAEs with various selection functions (i.e., limiting \ew\ and UV magnitudes).
Thus, 
it is crucial to investigate whether the selection functions of LAEs 
affect the \ew\ distribution results. 
This is important because previous observational studies have pointed out that 
fainter continuum objects have larger \ew\ values, the so-called Ando effect 
(e.g., \citealt{ando2006, stark2010, furusawa2016}).  
With our MUSE LAE sample, we examine the \ew\ distribution and its redshift evolution
between $z=2.9$ and $6.6$.

This paper is organized as follows. 
We describe our data and LAE sample in \S \ref{sec:data}.
In \S \ref{sec:uv_prop}, we 
derive UV continuum slopes ($\beta$) 
and UV absolute magnitudes (\muv) of our LAEs. 
In this section, a correlation between \muv\ and $\beta$ 
and the redshift evolution of $\beta$ are presented. 
In \S \ref{sec:flya_measurement}, 
we derive \lya\ fluxes 
based on the curve of growth technique 
and examine AGN activity of our LAE sample in \S \ref{sec:AGN}. 
In \S \ref{sec:lya_ew}. 
we show the \ew\ distribution and its redshift evolution. 
The Ando effect is examined in \S \ref{sec:ando_effect},
followed by properties of very large \ew\ LAEs in \S \ref{sec:large_ew}. 
Discussion in the context of \ew\ and comparisons between observations and theoretical studies 
are presented in \S \ref{sec:discussion}, 
and  our summary and conclusions are presented in \S \ref{sec:summary}.
Throughout this paper, magnitudes are given in the AB system
\citep{oke1983} and we assume a $\Lambda$CDM cosmology 
with $\Omega_{\small m} = 0.3$, $\Omega_{\small \Lambda} = 0.7$
and $H_{\small 0} = 70$ km s$^{-1}$ Mpc$^{-1}$.

\section{Data and sample}
\label{sec:data}

\subsection{Spectroscopy with MUSE} \label{subsec:obs}

We carried out observations with MUSE in the UDF 
between September $2014$ and February $2016$  
under the MUSE consortium GTO (PI: R. Bacon). 
The wavelength range of MUSE is $4750-9300$ \AA 
and the typical instrumental spectral resolution is $R \sim 3000$. 
\cite{Bacon2017} (hereafter B17) provide more details about the observations and data reduction. 
Briefly, the UDF was observed with MUSE in two different integration times (see Figure 1 in B17). 
The \mosaic\ field is the medium deep region consisting of nine pointings of 
$1$ arcmin$^{2}$ ($9$ arcmin$^{2}$ in total). 
In this region, each pointing has a $10$ hours exposure time. 
The \udft\ field is the ultra deep region, covering $1$ arcmin$^{2}$. 
In this region, the total exposure time is $31$ hours. 
The spatial scale is $0''.2 \times 0''.2$ per spatial pixel 
and the spectral sampling is $1.25$ \AA\ per spectral pixel. 

\subsection{Source extractions} \label{subsec:source_extractions}

The source extraction of objects and the construction of the parent catalog 
are given in B17 and \cite{Inami2017} (hereafter I17). 
In short, objects were detected and extracted using two methods. 

The first method uses the catalog of \cite{rafelski2015} as a positional prior. 
In \cite{rafelski2015}, photometry has been performed for $9927$ objects in the UDF 
with the latest and the deepest HST data 
covering the wavelength ranges from far ultraviolet (FUV) to near-infrared (NIR).
Using the sky coordinates of each object from the catalog of \cite{rafelski2015},  
we searched for spectral features (absorption or emission lines). 

The second method is based on our custom made software 
{\tt ORIGIN} (Mary et al. in preparation). 
{\tt ORIGIN} blindly searches for emission line objects (see B17 for the detail). 
The strength of {\tt ORIGIN} is that we can detect emission line objects  
without HST images as positional priors. 
The {\tt ORIGIN}-only objects without HST counterparts 
are candidates for very large \ew\ LAEs. 
This is because non-detections of HST images indicate that 
their continuum fluxes are extremely faint, increasing their \ew. 
These objects are presented in B17 and their properties will be presented elsewhere. 

\subsection{Parent \lya\ emitters
 sample} \label{subsec:parent_sample}

The parent LAE sample was constructed by I17 
with the following two criteria: 

\begin{itemize}

\item 
We selected LAEs with secure redshifts 
$2.9 < z < 6.6$ (`TYPE=6' and `CONFID=2 and 3'). 
 
\item 
As we describe in detail in \S \ref{sec:flya_measurement}, 
we created continuum-subtracted narrowband images of \lya\ emission 
in the same way as in \cite{drake2016} and \cite{Drake2017} (hereafter D17). 
Based on the narrowband images, we estimated \lya\ fluxes and errors (see \S \ref{subsec:measurement_llya}). 
We imposed a minimum signal-to-noise ratio (S/N) in \lya\ flux of $5$. 
The minimum S/N adopted in the present study is slightly lower than 
the S/N $=6$ used in \cite{Leclercq2017} (hereafter L17). 
The higher S/N limit is important in L17 
because their goal is to detect diffuse faint \lya\ emission on an individual basis. 
In this study, we chose the S/N cut of $5$ to increase the number of LAEs. 

\end{itemize}

A fraction of LAEs in the \udft\ field are also detected in the \mosaic\ field. 
In these overlapped cases, 
we adopted the results in the \udft\ field because this field is deeper than the \mosaic\ in \lya. 
After removing those overlapped objects, 
there are $156$ and $526$ parent LAEs in the \udft\ and \mosaic\ fields, respectively. 

For these objects, we performed visual inspection. 
In this procedure, we first removed spurious objects 
\footnote{
These include LAEs with OH sky line contamination 
and with the noisy \lya\ lines. 
}
and next removed LAEs with close companion LAEs  
whose individual \lya\ fluxes are affected by the companions' \lya\ fluxes. 
In total, $11$ objects were removed from the sample.

\subsection{Our \lya\ emitters selected with MUSE and public HST data} \label{subsec:public_hst}

For robust estimates of \ew, 
it is important to obtain accurate continuum fluxes at $1216$ \AA. 
As can be seen in Figure $9$ of \cite{bacon2015} and in Figure $12$ of B17,  despite the high sensitivity of MUSE, 
it is difficult to precisely determine continuum fluxes for faint objects. 

Therefore,  we used the public HST photometry catalog of \cite{rafelski2015}. 
We describe the HST data in \S \ref{subsubsec:public_hst} and 
then construct our final  LAE sample in \S \ref{subsubsec:our_sample}.

\subsubsection{Public HST data} \label{subsubsec:public_hst}

The catalog of \cite{rafelski2015} is the same as the catalog 
we used as a positional prior for source extractions (\S \ref{subsec:source_extractions}).  
At $z\sim2.9-6.6$, the rest-frame FUV continuum roughly 
corresponds to $8000-16000$ \AA\ in the observed frame. 
Thus, we used the public HST data from F775W to F160W 
depending on the redshifts of the objects. 
Table \ref{tab:hst_data} summarizes the public HST photometry data used in this study. 

For the objects detected with the positional priors, 
we used total magnitudes from \cite{rafelski2015}.  
The total magnitudes were obtained from the Kron radius (\citealt{kron1980}) 
and were carefully corrected for aperture-matched PSFs and Galactic extinction. 
For the objects detected only by {\tt ORIGIN}, 
we performed our own photometric analysis using {\tt NoiseChisel} developed by 
\cite{akhlaghi2015} (see B17 for the procedure).

\subsubsection{Our \lya\ emitters sample} \label{subsubsec:our_sample}

One has to take the PSF difference into account to fairly compare HST data with MUSE data. 
As described in B17 and I17, 
the segmentation maps of MUSE data cubes were 
based on the segmentation map of  HST data (\citealt{rafelski2015}) 
convolved with the MUSE PSF, typically $FWHM \approx 0''.6$ (see the top panel of Figure $7$ in B17). 
The B17 and I17 works carefully assigned each MUSE-detected object to an HST counterpart. 
To do so, B17 and I17 examined the narrowband images. 
In this procedure, 
$78$ LAEs were found to have more than one HST counterparts. 
These objects were removed from our sample to obtain a clean sample. 
For the rest of the sample with a single HST counterpart, 
we could directly compare  MUSE-based \lya\ fluxes with HST-based continuum fluxes.

As we describe in detail in \S \ref{sec:uv_prop}, 
we used two or three HST wave bands to derive UV continuum slopes. 
Therefore, we also applied the following HST detection criterion to our LAEs: 
 At least two HST bands are detected above $2 \sigma$. 
The typical $2\sigma$ limiting magnitudes within $0''.2$ radius apertures 
correspond to apparent magnitudes of $29.2-31.1$ (see Table \ref{tab:hst_data}).

After imposing this criterion on our objects, 
we are left with $80$ and $337$ LAEs in the \udft\ and \mosaic\ fields, respectively. 
The redshift distribution of the two fields are shown in the left panel of Figure \ref{fig:beta_muv_distribution}. 
For the remainder of the present paper, 
we use the sample with HST detections above $2\sigma$. 
Table \ref{tab:lae_sample} summarizes our LAE sample.

We discuss possible bias effects due to our selection technique 
in \S \ref{subsec:limitations}.

\begin{table*}
\centering
\caption{Summary of Our LAE sample.} 
\begin{tabular}{cccccc}
\hline
Field & $N_{\rm tot}$  & $N_{\rm analyzed}$ & & & \\ 
 &  &  & $<z>=3.6$ & $<z>=4.9$ & $<z>=6.0$ \\
\hline
\udft\ & $156$ & $80$  & $56$ & $18$ & $6$  \\
\mosaic\ & $526$ & $337$ & $224$ & $90$  & $23$  \\
\hline 
Total & $682$ & $417$  & $280$ & $108$ & $29$  \\
\hline
\end{tabular}
\tablefoot{
$N_{\rm tot}$ denotes the total number of spectroscopic LAEs in I17 
that have secure redshifts and \lya\ flux S/N $>5.0$. 
$N_{\rm analyzed}$ is the number of LAEs analyzed in this paper. 
Numbers denote samples with HST detections above 
$2\sigma$ in HST wave bands listed in Table \ref{tab:hst_passbands}.  
}
\label{tab:lae_sample}
\end{table*}


\section{Ultraviolet continuum properties obtained with HST} \label{sec:uv_prop}

\subsection{Ultraviolet magnitudes and continuum slopes} \label{sec:uv_prop_method}

Ultraviolet continuum slopes are estimated by fitting two or three HST magnitudes. 
From the definition of UV continuum slopes, $f_{\rm \lambda} \propto \lambda^{\beta}$, 
the relation between AB magnitudes and wavelengths in \AA\ is expressed as 
\begin{eqnarray}
\label{eq:beta}
{\rm mag} = -2.5 {\rm log} (\lambda^{\beta+2}) + A, 
\end{eqnarray}
where $A$ is a constant corresponding to the amplitude. 
We chose passbands so that 
Ly$\alpha$ emission or intergalactic medium (IGM) absorption do not affect the photometry. 
In order to calculate $\beta$ values as uniform as possible 
at rest-frame wavelengths, 
we divided our LAEs into three redshift bins
based on their spectroscopic redshifts, $z_{\rm sp}$: 
$2.90 \leqq z_{\rm sp} \leqq 4.44$, 
$4.44 < z_{\rm sp} \leqq 5.58$, 
and 
$5.58 < z_{\rm sp} \leqq 6.66$, 
with mean redshifts of $z=3.6$, $4.9$, and $6.0$, respectively. 
The number of LAEs in each redshift bin are listed in Table \ref{tab:lae_sample}, 
and the relevant HST filters are listed in Table \ref{tab:hst_passbands}. 
With the typical wavelengths of the filters, 
our $\beta$ values probe UV continuum slopes 
in the  rest-frame wavelength ranges of $\sim1700-2400$ \AA, 
which are consistent with those in \cite{bouwens2009}: $1600-2300$ \AA.
Typically we used three filter bands to determine $\beta$. 
However, owing to the limited spatial coverage of F140W, 
the determination of $\beta$ rely on the remaining two filters for some objects.  
We checked and confirmed that the $\beta$ measurements 
are not statistically affected by the lack of F140W 
\footnote{
The lack of F140W can affect the results at $z\sim4.9$ and $6.0$ (see Table \ref{tab:hst_passbands}). 
Basically, most \udft\ LAEs are in the coverage of F140W. 
Thus, using these LAEs,  we derive two $\beta$ values: with and without F140W. 
To evaluate the effect, we performed the Kormogorov-Smirnov (K-S) test for the two $\beta$ distributions. 
We obtain the $p$ values of $0.36$ and $0.99$ for $z\sim4.9$ and $6.6$, respectively, 
indicating that the $\beta$ distributions cannot be distinguished from each other. 
However, the uncertainties in $\beta$ measurements become smaller if we include F140W. 
}. 

With $\beta$ and $A$ values in equation (\ref{eq:beta}), 
we estimate apparent magnitudes at $1500$ \AA, $m_{\rm 1500}$, as follows: 
\begin{eqnarray}
\label{eq:app_m}
m_{\rm 1500} 
= -2.5 {\rm log} (\bigl\{1500\times(1+z_{\rm sp})\bigr\}^{\beta+2}) + A. 
\end{eqnarray}
From $m_{\rm 1500}$, we obtain \muv\ as 
\begin{eqnarray}
M_{\rm UV} = m_{1500} -5 {\rm log}(d_{L}/{\rm 10pc}) + 2.5 {\rm log}(1+z_{\rm sp}), 
\end{eqnarray}
where $d_{L}$ indicates the luminosity distance in parsec (pc) corresponding to 
the spectroscopic redshift, $z_{\rm sp}$, derived in I17.

We estimate apparent magnitudes at $1216$ \AA, $m_{\rm 1216}$, 
as in equation (\ref{eq:app_m}). 
Using $m_{\rm 1216}$, we obtain continuum fluxes at $1216$ \AA\ 
in erg cm$^{-2}$ s$^{-1}$ Hz$^{-1}$, $f_{\rm \nu, cont}$, from the relation 
\begin{eqnarray}
\label{eq:f_nu}
f_{\rm \nu, cont} = 10^{-0.4(m_{\rm 1216}+48.6)}.
\end{eqnarray}
Finally, we derive $f_{\rm \lambda, cont}$ from $f_{\rm \nu, cont}$ 
as follows: 
\begin{eqnarray}
\label{eq:f_lam}
f_{\rm \lambda, cont} = 
f_{\rm \nu, cont} \times \frac{c}{\bigl\{ 1216(1+z_{\rm sp})\bigr\}^{2}}, 
\end{eqnarray}
where $c$ is the speed of light in \AA\ s$^{-1}$. 

To estimate the physical quantities and their errors, 
we applied a Monte Carlo technique as we describe below.
With HST magnitudes and their errors, we generated $300$ mock magnitudes 
for each passband listed in Table \ref{tab:hst_passbands} 
under the assumption that the magnitude distribution is a Gaussian.
We take the low-$z$ bin as an example. 
With $300$ sets of mock magnitudes, F775W, F850LP, and F105W, 
we derive $300$ sets of $\beta$ and $A$ values with equation (\ref{eq:beta}).
We then obtain $300$ sets of \muv\ and $f_{\rm \lambda, cont}$
from equations (\ref{eq:app_m}) $-$ (\ref{eq:f_lam}). 
The median and standard deviation of the distribution of measurements 
are adopted as the measured and error values, respectively.

\begin{figure*}[tbp]
\centering
\includegraphics[width=20cm]{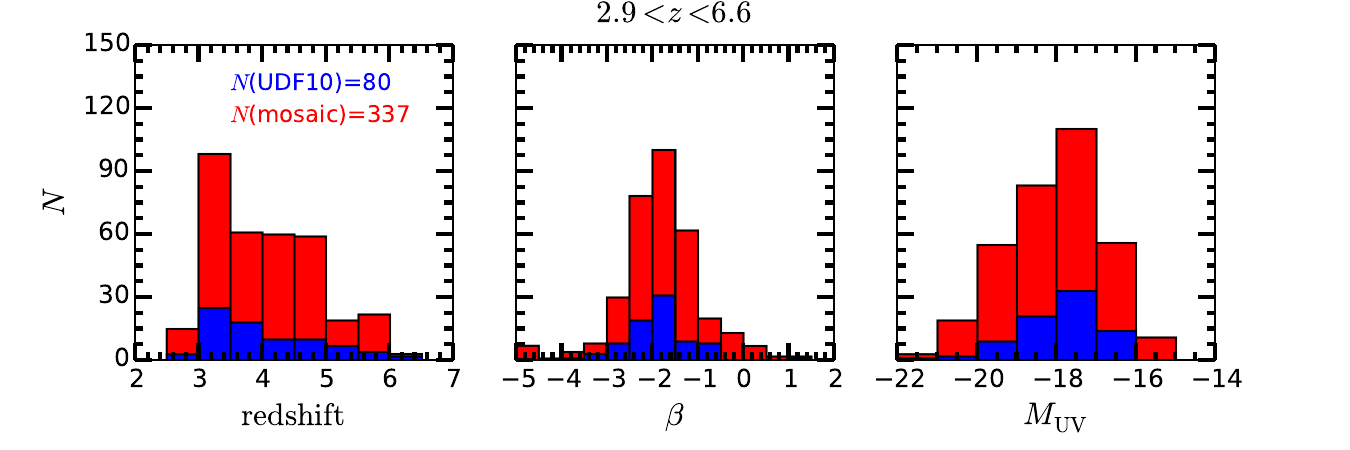}
\caption[]
{
Left, middle, and right panels: Distributions of $z$, $\beta$ and \muv\ for the entire sample at $2.9 < z < 6.6$, respectively. 
In each panel, the blue and red histograms correspond to the distributions 
for \udft\ and \mosaic, respectively. 
A two-sample Kolmogorov-Smirnov test (K-S test) results in 
the $p$-value of $0.84$, $0.25$, and $0.32$ for the 
two $z$, $\beta$ and \muv\ distributions, respectively, 
indicating that the distributions of the values in the two fields cannot be distinguished 
from each other.
}
\label{fig:beta_muv_distribution}
\end{figure*}

The middle panel of Figure \ref{fig:beta_muv_distribution} shows the $\beta$ distribution 
for the entire sample of LAEs. The $\beta$ values range from $-5$ to $1$ with a median value of $-1.81$. 
The values $\beta \lesssim -3$ are physically unlikely (e.g., \citealt{schaerer2003}). 
We find that objects with very steep values, for example, $\beta \lesssim -3$, 
have uncertainties on $\beta$ as large as $1.0$. 
For the combined sample of LAEs in the \udft\ and \mosaic\ fields, 
we calculated the mean, median, standard deviation, and standard error values 
for each redshift bin. 
These values are listed in Table \ref{tab:quantities}. 

The right panel of Figure \ref{fig:beta_muv_distribution} shows 
the \muv\ distribution for our LAEs. 
The median value, $-17.9$, 
is more than two orders of magnitude fainter than 
previous high $z$ LAE studies based on the narrowband technique 
(\citealt{shimasaku2006, ouchi2008}) 
and spectroscopy (\citealt{stark2010, cassata2015}). 
The typical \muv\ value in these studies is roughly $-20.5$. 
In our LAE sample selection, we included all objects with 
HST detections above $2\sigma$ in multiple wave bands. 
The corresponding lowest \muv\ values are $\sim-16$, $-17$, and $-18$ 
at $z\sim3.6$, $4.9$, and $6.6$, respectively.

\begin{table*}
\centering
\caption{Wave bands used to derive the $UV$ continuum slope for individual galaxies.}
\begin{tabular}{cccc}
\hline
Redshift &  Mean  & Filters  & Rest-frame \\
range & redshift &  & wavelengths (\AA) \\
(1) & (2) & (3) & (4) \\
\hline
\object{$2.90 \leqq z_{\rm sp} \leqq 4.44$}  & 3.6 & F775W, F850LP, F105W & $1700-2300$ \\ 
\object{$4.44 < z_{\rm sp} \leqq 5.58$} & 4.9 & F105W, F125W, F140W$^{a}$ & $1800-2100$ ($1800-2400$)$^{b}$ \\ 
\object{$5.58 < z_{\rm sp} \leqq 6.66$} & 6.0 & F125W, F140W$^{a}$, F160W & $1800-2200$ \\
\hline
\end{tabular}
\tablefoot{
(1) Spectroscopic redshift ranges of the three redshift bins.
(2) Mean redshift of each redshift bin.
(3) HST filters used to estimate UV continuum slopes.   
(4) Typical rest-frame wavelengths probed by  UV continuum slopes.\\
$^{a}$ F140W is used if it is available. \\
$^{b}$ Value in the parenthesis is the wavelength range 
in the case that F140W is available.
}
\label{tab:hst_passbands}
\end{table*}

\begin{table*}
\centering
\caption{Summary of physical quantities.} 
\begin{tabular}{ccccccc}
\hline
Quantity & $z$ & $N$ & mean & median & $\sigma$ & $\sigma/\sqrt{N}$ \\ 
(1) & (2) & (3) & (4) &  (5) & (6) & (7) \\
\hline
$\beta$ & $3.6$ & $280$ & $-1.62$ & $-1.73$ & $0.72$ & $0.04$ \\ 
            & $4.9$ & $108$ & $-2.17$ & $-2.22$ & $1.57$ & $0.15$ \\  
            & $6.0$ & $29$ & $-2.10$ & $-2.321$ & $1.05$ & $0.19$ \\  
\hline     
\muv\    & $3.6$ & $280$ & $-17.7$ & $-17.6$ & $1.1$ & $0.1$ \\ 
            & $4.9$ & $108$ & $-18.4$ & $-18.4$ & $1.0$ & $0.1$ \\ 
            & $6.0$ & $29$ & $-19.1$ & $-19.0$ & $1.1$ & $0.2$ \\  
\hline     
\llya\    & $3.6$ & $280$ & $41.9$ & $41.9$ & $0.4$ & $0.1$ \\ 
            & $4.9$ & $108$ & $42.1$ & $42.0$ & $0.4$ & $0.1$ \\
            & $6.0$ & $29$ & $42.5$ & $42.5$ & $0.4$ & $0.1$ \\ 
\hline     
\ew\     & $3.6$ & $280$ & $113$ & $87$ & $96$ & $6$ \\
           & $4.9$ & $108$ & $83$ & $57$ & $88$ & $8$ \\ 
           & $6.0$ & $29$ & $130$ & $97$ & $120$ & $22$ \\ 
\hline
\end{tabular}
\tablefoot{
(1) Physical quantity;
(2) redshift of the sample; 
(3) number of objects; 
(4)-(7) mean, median, standard deviation, and standard error values.
}
\label{tab:quantities}
\end{table*}

\subsection{Correlation between \muv\ and $\beta$} \label{subsec:muv_beta}

For dropout galaxies, a uniform picture has emerged that  
$\beta$ values become steeper at fainter \muv\ at various redshifts from $z\sim1$ to $8$ 
(e.g., \citealt{bouwens2009, wilkins2011, bouwens2012, bouwens2014, kurczynski2014}).  
While \cite{finkelstein2012, dunlop2012, hathi2016} claimed that the correlation is not clear, 
\cite{kurczynski2014, bouwens2014, rogers2014} showed that 
the discrepant results are due to systematics and biases. 
Once corrected for these systematics and biases, 
the slope is consistently $d\beta$/d\muv\   $\approx -0.10$. 
Since $\beta$ values become steeper if the dust content is low (\citealt{meurer1999}), 
this anti-correlation is interpreted as fainter \muv\ galaxies having lower dust contents. 

Several previous studies examined $\beta$ in LAEs at $3<z<7$ 
(e.g., \citealt{ouchi2008, ono2010b, stark2010, jiang2013}). 
However, compared to the typical magnitude range of the dropout galaxies, $-22 <$ \muv\ $< -15$, 
the magnitude range in the LAE studies is narrow, $-22 <$ \muv\ $< -19$. 
Because our LAEs have a UV magnitude range that is
comparable to that for dropout galaxies, $-22 <$ \muv\ $< -16$, 
we compared our $\beta$ values with those of dropout galaxies. 

\begin{figure*}[t]
\centering
\includegraphics[width=20cm]{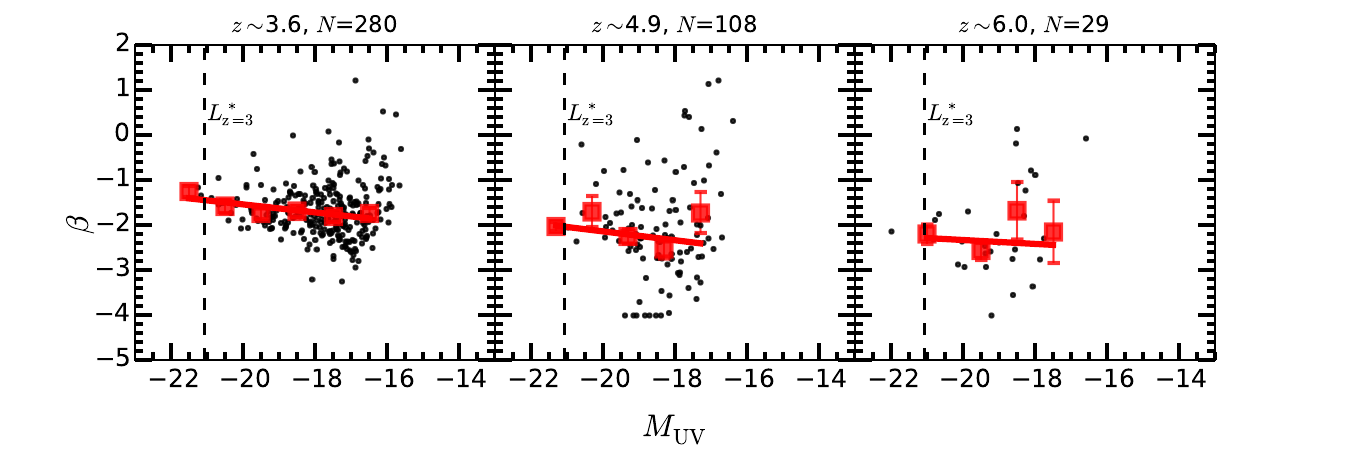}
\caption[]
{
From left to right, $\beta$ plotted against \muv\ for $z\sim3.6$, $4.9$, and $6.0$. 
The small black circles indicate individual LAEs. 
The vertical dashed line indicates the characteristic UV luminosity at $z\sim3$, 
$L^{*}_{\rm z=3} = -21.07$ (\citealt{steidel1999}). 
The red squares show biweight mean values of $\beta$ at each \muv\ bin. 
The biweight mean is a robust statistic for determining the central location of a distribution. 
The standard deviation of the biweight mean is determined 
based on bootstrap simulations at each magnitude bin. 
The solid red line is the best-fit linear function to the biweight mean values. 
The slopes are $d\beta$/d\muv\   $= -0.09\pm0.03$, $-0.10\pm0.06$, and $-0.04\pm0.15$
for $z\sim3.6$, $4.9$, and $6.0$, respectively. 
}
\label{fig:MUV_beta}
\end{figure*}

Figure \ref{fig:MUV_beta} plots $\beta$ against \muv\ for our individual LAEs. 
To quantify the relation, we calculated the biweight mean of $\beta$ at each magnitude bin
(cf. \citealt{bouwens2012, bouwens2014}). 
The biweight mean and error values are listed in Table \ref{tab:biweight_sample}. 
We fit the biweight mean values with a linear function. 
The slopes are $d\beta$/d\muv\   $= -0.09\pm0.03$, $-0.10\pm0.06$, and $-0.04\pm0.15$
for $z\sim3.6$, $4.9$, and $6.0$, respectively. 
From Figure \ref{fig:MUV_beta}, 
we see that $\beta$ values become steeper at fainter \muv, 
in agreement with the previous findings of \cite{bouwens2012}. 

\begin{figure*}[t]
\centering
\includegraphics[width=10cm]{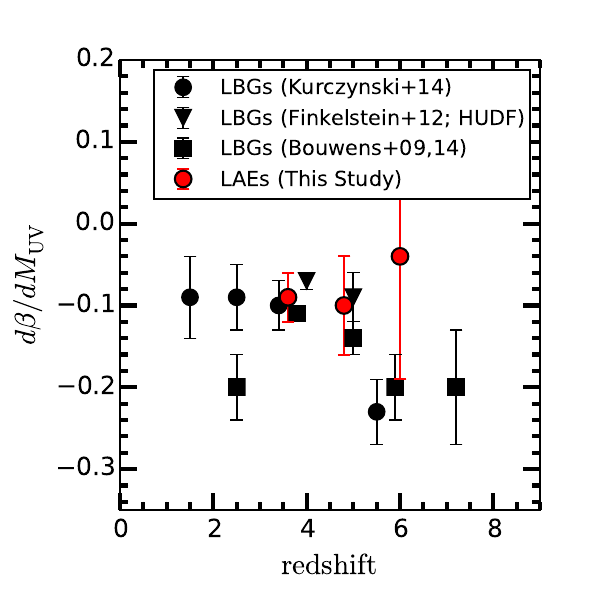}
\caption[]
{
Derivative of $\beta$ with UV magnitude plotted against redshift, $z$. 
Our LAEs, denoted as red circles, are placed at mean redshifts $z\sim3.6$, $4.9$, and $6.6$. 
}
\label{fig:z_dbdM}
\end{figure*}

In Figure \ref{fig:z_dbdM}, we compare our $d\beta$/d\muv\ values 
with those of dropout galaxies (\citealt{bouwens2009, bouwens2014, finkelstein2012, kurczynski2014}). 
We find that our $d\beta$/d\muv\ of LAEs are in good agreement with previous studies of dropout galaxies. 
These results therefore indicate that fainter UV continuum LAEs have lower dust contents.

\begin{table*}
\centering
\caption{Biweight mean of physical quantities as a function of ultraviolet luminosity.} 
\begin{tabular}{ccccc}
\hline
\muv & log \llya & $\beta$ & \ew & $N$ \\ 
(1) & (2) & (3) & (4) &  (5)  \\
\hline
& & $z\sim3.6$ & & \\
\hline
$-21.5$ & $42.0\pm0.7$ & $-1.26\pm0.07$ & $32\pm14$ & $2$ \\
$-20.5$ & $42.4\pm0.3$ & $-1.58\pm0.10$ & $23\pm10$ & $11$ \\
$-19.5$ & $42.3\pm0.1$ & $-1.74\pm0.07$ & $44\pm6$ & $29$ \\
$-18.5$ & $42.1\pm0.1$ & $-1.69\pm0.07$ & $65\pm7$ & $57$ \\
$-17.5$ & $41.9\pm0.1$ & $-1.80\pm0.06$ & $90\pm6$ & $107$ \\
$-16.5$ & $41.7\pm0.1$ & $-1.74\pm0.07$ & $140\pm12$ & $63$ \\
\hline     
& & $z\sim4.9$ & & \\
\hline
$-21.5$ & $43.3\pm0.1$ & $-2.02\pm0.01$ & $85\pm19$ & $2$ \\
$-20.5$ & $42.4\pm0.1$ & $-1.70\pm0.33$ & $32\pm9$ & $9$ \\
$-19.5$ & $42.3\pm0.1$ & $-2.24\pm0.12$ & $47\pm8$ & $31$ \\
$-18.5$ & $42.0\pm0.1$ & $-2.53\pm0.19$ & $46\pm8$ & $35$ \\
$-17.5$ & $41.8\pm0.1$ & $-1.72\pm0.44$ & $78\pm16$ &$27$ \\
\hline     
& & $z\sim6.0$ & & \\
\hline
$-21.0$ & $42.6\pm0.2$ & $-2.19\pm0.20$ & $24\pm8$ & $6$ \\
$-19.5$ & $42.6\pm0.2$ & $-2.55\pm0.19$ & $91\pm39$ & $9$ \\
$-18.5$ & $42.5\pm0.1$ & $-1.68\pm0.63$ & $173\pm49$ & $10$ \\
$-17.5$ & $42.4\pm0.2$ & $-2.15\pm0.68$ & $155\pm134$ & $3$ \\
\hline
\end{tabular}
\tablefoot{
The uncertainty values are the standard errors 
derived based on bootstrap simulations. 
The values represent how the median values are well constrained. 
}
\label{tab:biweight_sample}
\end{table*}


\subsection{Redshift evolution of $\beta$} \label{subsec:evolution_beta}

\begin{figure*}[tbp]
\centering
\includegraphics[width=18cm]{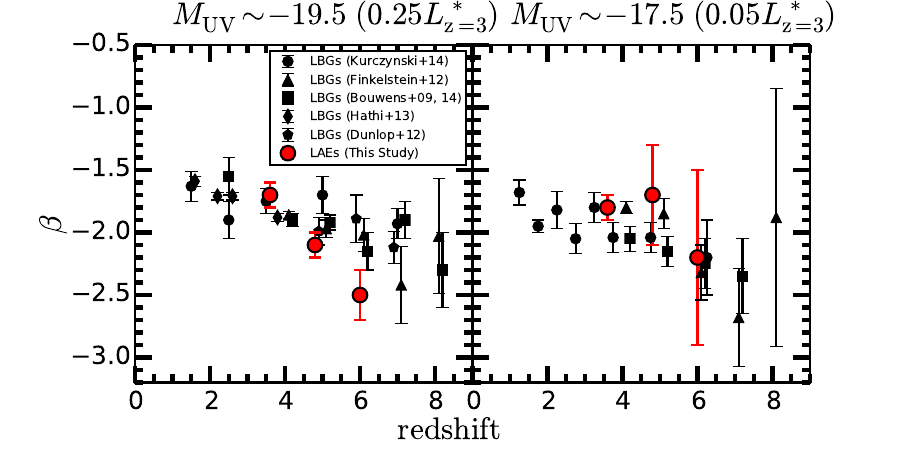}
\caption[]
{
Left and right panels: Redshift evolution of $\beta$ values 
at bright (\muv\ $\sim-19.5$) and faint (\muv\ $\sim-17.5$) 
UV absolute magnitudes, respectively. 
}
\label{fig:z_beta}
\end{figure*}

Previous studies on continuum-selected galaxies have shown that $\beta$ values become steep at high $z$ 
(\citealt{bouwens2009, dunlop2012, finkelstein2012, hathi2013, bouwens2014, kurczynski2014}). 
Since we derived $\beta$ values in a uniform manner at $2.9 < z < 6.6$, 
it is interesting to see if LAEs have a similar redshift evolution in $\beta$. 
Figure \ref{fig:z_beta} shows the redshift evolution of $\beta$ for our LAEs. 
We also include data points of dropout galaxies in the literature mentioned above. 
To perform fair comparisons of $\beta$ at various redshifts, 
we investigated $\beta$ evolutions in two \muv\ bins, $\sim-19.5$ and $-17.5$. 
These \muv\ values correspond to $0.25$ and $0.05$ $L^{*}_{\rm z=3}$, respectively, 
where $L^{*}_{\rm z=3}$ is $-21.07$ (\citealt{steidel1999}). 
We chose these \muv\ values to compare our results with those in \cite{kurczynski2014} 
who used the same \muv\ bins. 

There are two results in Figure \ref{fig:z_beta}. 
First, we find that our $\beta$ values are consistent with those in dropouts 
within $1\sigma$ uncertainties at a given \muv. 
At first glance, the result is at odds with the result of \cite{stark2010}. 
These authors found that dropout galaxies with \lya\ emission have 
steeper $\beta$ compared with those without \lya\ emission 
at the UV magnitude range from $-21.5$ to $-20.0$. 
However, as can be seen from Figure $14$ in \cite{stark2010}, 
the $\beta$ difference becomes negligible in their faintest bin, \muv$= -20.0$. 
Therefore, given the very faint \muv\ of our LAEs (see Figure \ref{fig:beta_muv_distribution}), 
it is not surprising that our LAEs and dropout galaxies have similar $\beta$.  
Second, we see a trend that $\beta$ becomes steeper at higher $z$ in LAEs, at least at bright \muv. 
This trend is also consistent with that in dropouts, 
indicating that the dust contents of LAEs is low at high $z$.

To summarize this section, 
we  presented UV continuum properties of our LAEs, 
which cover a wide range of \muv. 
We demonstrated that $\beta$ values in LAEs are 
in good agreement with those in dropout galaxies 
at a given redshift or \muv. 
The results indicate that dust contents become smaller  
for higher $z$ and fainter \muv\ galaxies.

\section{Accurate \lya\ fluxes obtained with MUSE} \label{sec:flya_measurement}

\subsection{Measurements of \lya\ fluxes} \label{subsec:measurement_llya}

\cite{wisotzki2016} and L17 have shown that \lya\ emission is significantly extended 
compared with UV emission 
not only statistically but also for individual objects.
To capture the extended \lya\ flux,  we adopted the curve of growth technique 
in the same manner as in \cite{wisotzki2016, drake2016, Drake2017, Leclercq2017}. \ The detailed procedure is provided in \S $3$ of D17. 
Briefly, we performed photometry on the \lya\ narrowband images 
after subtracting the local background and masking out nearby objects. 
We applied various sizes of annuli until 
the curve of growth reaches the background level. 
The cumulative flux is adopted as the total \lya\ flux, 
while the error flux is estimated from the variance cube.

We note that 
our \lya\ fluxes are not corrected for the Galactic extinction. 
However, correction factors would be very small in the UDF as we describe below. 
In the UDF, \cite{rafelski2015} have investigated the Galactic extinction. 
In the F606W and F775W bands, whose wavelengths coverage matches 
those of our \lya\ lines, the Galactic extinction values are $0.023$ and $0.016$, respectively. 
These differences in magnitudes correspond to $\sim2\%$ differences in fluxes. 
Therefore, regardless of the correction for the Galactic extinction, 
our results remain unchanged.

\begin{figure*}[tbp]
\centering
\includegraphics[width=20cm]{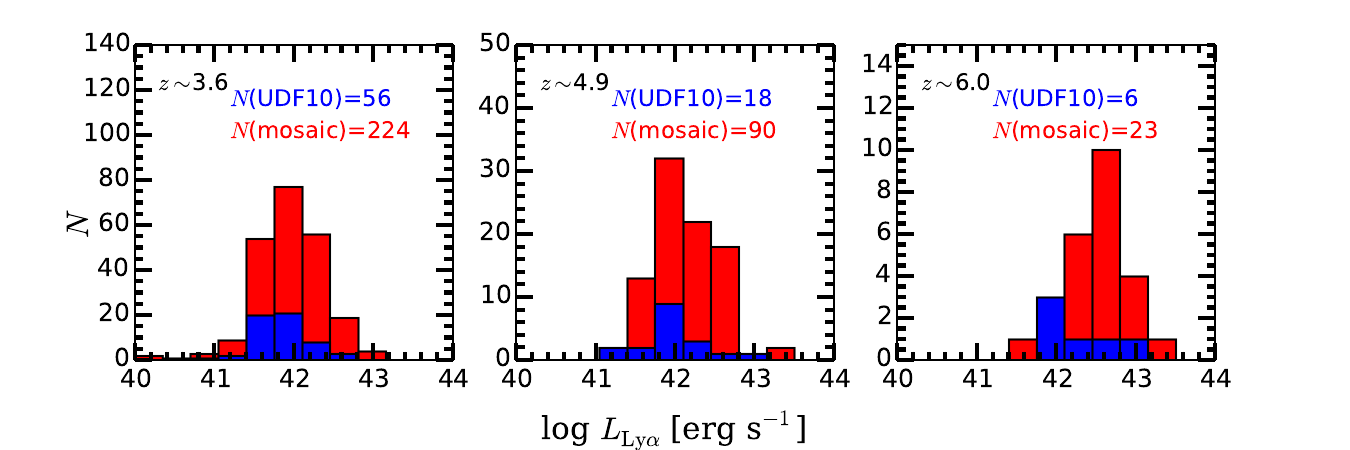}
\caption[]
{
From left to right,  \lya\ luminosity distributions for $z\sim3.6$, $4.9$, and $6.0$. 
The blue and red histograms correspond to the distributions for \udft\ and \mosaic, respectively. 
Two sample  K-S tests result in $p$ values of $0.01$, $0.34$, 
and $0.08$ for $z\sim3.6$, $4.9$, and $6.0$, respectively, 
indicating that the distributions of \llya\ values in the two fields are statistically 
different from one another at least at $z\sim3.6$. 
}
\label{fig:LLyA_distribution}
\end{figure*}

Figure \ref{fig:LLyA_distribution} shows the distribution of \lya\ luminosities, \llya, for our LAEs. The 
\llya\ values span the range from log \llyaunil\ $\approx 41.0$ to 43.0. 
Because we obtained deeper data in \udft\ than in \mosaic, 
we investigated the \lya\ depth difference in the two fields. 
We found that the mean \lya\ flux in \udft\ is $1.3$, $1.3$, and $2.0$ times fainter than in \mosaic\ 
at $z\sim3.6$, $4.9$, and $6.0$, respectively
\footnote
{
These correspond to the log \llyaunil\ difference of $0.1$, $0.1$, and $0.3$ 
at $z\sim3.6$, $4.9$, and $6.0$, respectively. 
}. 
The mean, median, standard deviation, and standard error values 
for the entire sample are listed in Table \ref{tab:quantities}.

\subsection{\muv\ and \llya} \label{subsec:muv_llya}

\begin{figure*}[tbp]
\centering
\includegraphics[width=20cm]{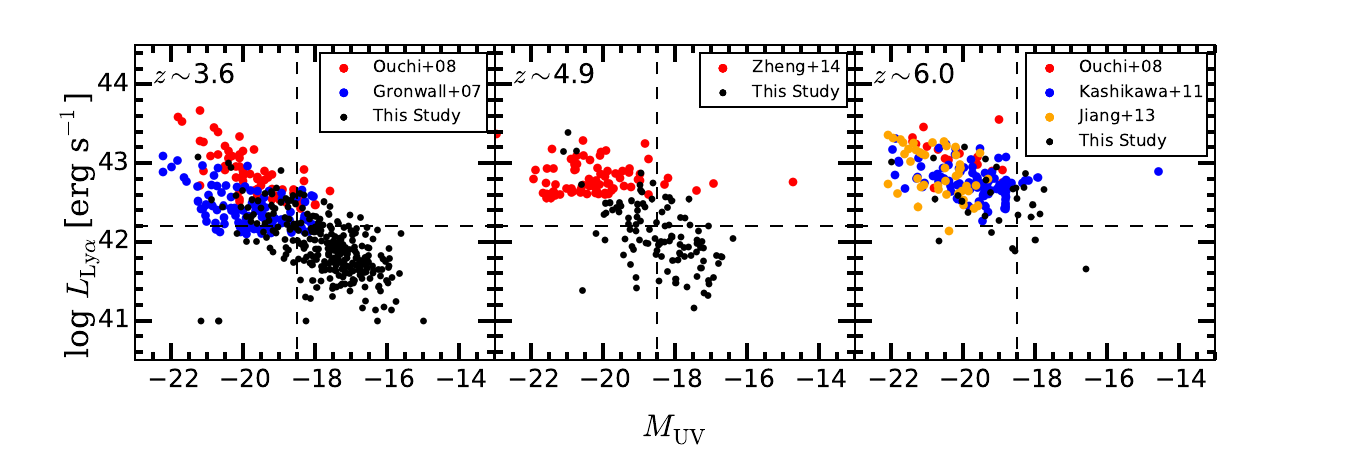}
\caption[]
{
From left to right, log \llya\ is plotted against \muv\ 
for $z\sim3.6$, $4.9$, and $6.0$. 
The black circles indicate our individual LAEs.
In each panel, objects with log \llyaunil\ $< 41.0$ 
are placed at $41.0$ for display purposes. 
\textit{Left panel:}
Red circles show spectroscopically confirmed LAEs from \cite{ouchi2008} 
at $z\sim3.1$ and $3.7$, 
while blue circles indicate a photometric LAE sample from \cite{gronwall2007}. 
\textit{Middle panel:}
Red circles correspond to $z\sim4.5$ LAEs studied by \cite{zhenya_zheng2014}. 
\textit{Right panel:}
Red circles show spectroscopically confirmed LAEs from \cite{ouchi2008} at $z\sim5.7$. 
Blue circles indicate spectroscopically confirmed LAEs from \cite{kashikawa2011} at $z\sim5.7$ and $6.5$, 
while orange circles are spectroscopically confirmed LAEs from \cite{jiang2013} 
at $z\sim5.7$, $6.5$, and $7.0$. 
In each panel, 
the vertical dashed line at \muv\ $=-18.5$ and 
the horizontal dashed line at log \llyaunil\ $=42.2$ 
show the cuts used for fair comparisons of \ew\ scale lengths at $2.9<z<6.6$ 
(see \S \ref{subsec:evolution_ew}). 
}
\label{fig:MUV_LLyA_literature}
\end{figure*}

In order to demonstrate the power of MUSE 
and the uniqueness of our sample, 
we compare our \muv\ and \llya\ with those in the literature 
in Figure \ref{fig:MUV_LLyA_literature}. 
As can be seen from the figure, 
our LAEs are fainter in both \muv\ and \llya\ than those in previous studies. 
In particular, at $z\sim3.6$ and $4.9$, 
lower ends of continuum and \lya\ fluxes 
are about an order of magnitude fainter than previous studies. 
At $z\sim6.0$, the magnitude (luminosity) difference is small 
between this study and the literature. 
This would be due to the small statistics at $z\sim6.0$ 
and because strong sky fluxes prevent 
us from detecting faint objects at $z\sim6.0$ (see Figure 5 in D17). 

Figure \ref{fig:MUV_LLyA_literature} also shows 
that brighter \muv\ objects have larger \llya. 
This trend is expected because both \muv\ and \llya\ 
values increase with the star formation rates (see also \citealt{matthee2017}).

\section{AGN activity in the sample} \label{sec:AGN}

It is known that AGN activity can also generate \lya\ emission 
as a result of ionizing photon radiation from AGNs (e.g., \citealt{malhotra2002}). 
Based on X-ray emission and high-ionization state emission lines 
(e.g., {\sc Civ} $\lambda 1549$ and He{\sc ii} $\lambda 1640$), 
previous studies have shown that the AGN fraction among LAEs is as low as $0-2\%$ at $z>3$ 
(e.g., \citealt{malhotra2003, gawiser2006, ouchi2008}). 
If this is the case, we expect $0-10$ AGNs among the present sample. 
Since we are interested in LAEs whose \lya\ emission is powered by star formation activity, 
we need to remove AGN-like LAEs from the sample. 

To do so, we first compared the sky coordinates of our LAEs 
with those in a very deep (7 Ms) archival  X-ray catalog (\citealt{luo2017}). 
The X-ray catalog includes objects detected in up to three X-ray bands: 
$0.5-7.0$ keV, $0.5-2.0$ keV, and $2-7$ keV. 
The average flux limits close to the HUDF are 
$1.9 \times 10^{-17}$, $6.4 \times 10^{-18}$, and $2.7 \times 10^{-17}$ 
erg cm$^{-2}$ s$^{-1}$ in the three X-ray bands.  
Following the procedure in \cite{herenz2017}, 
a cross-matching is regarded as successful if an LAE has a counterpart within an aperture.
We  adopted the aperture size of three times the X-ray positional error, 
which is the same aperture size as adopted in \cite{herenz2017}. 
We found that an AGN-LAE: LAE (AGN) ID is $6565$ ($758$), 
where AGN ID is taken from \cite{luo2017}. 
The AGN has not been spectroscopically identified in previous searches 
for optical counterparts of AGNs. 
We listed the object in Table \ref{tab:AGN_LAEs} 
and removed it from the sample.

Secondly, we made use of \lya\ luminosities, \llya. 
Recently, \cite{konno2016} have examined \llya\ of LAEs at $z\sim2$. 
The authors have revealed that bright LAEs with log \llyaunil\ $>43.4$ have X-ray or radio counterparts. 
Thus, \cite{konno2016} have concluded that very bright LAEs at $z\sim2$ are AGNs. 
Based on this result, 
we regard an LAE to be an AGN if log $L$(\lya/\unil) $>43.4$. 
None of our LAEs satisfy this criterion. 

Finally, we assessed the full width half maxima (FWHM) of \lya\ spectral lines
in the catalog presented in I17. 
It is expected that Type $1$ AGNs have broad \lya\ emission lines. 
None of our LAEs have FWHM values larger than $1000$ km s$^{-1}$. 

We conclude that there is at least one obvious Type $1$ AGN in our LAE sample. 
In addition, hidden Type $2$ AGNs may present among the sample.

\begin{table*}
\centering
\caption{Properties of a X-ray detected AGN-like LAE.} 
\begin{tabular}{cccccccl}
\hline
MUSE ID & Chandra 7Ms ID& $z$ & \ew  & log \llya & \muv  & $\beta$ & FWHM(\lya)\\ 
(1) & (2) & (3) & (4) &  (5) & (6) & (7) & (8)\\
\hline
$6565$ & $758$ & $3.20$ & $132\pm116$ & $41.6\pm0.4$ & $-16.4\pm0.2$ & $-1.9\pm0.4$ & $209\pm15$ \\
\hline
\end{tabular}
\tablefoot{
ID and physical quantities of an AGN-LAE whose optical counterpart has not been identified in previous studies. 
Chandra 7 Ms IDs are taken from \cite{luo2017}. 
}
\label{tab:AGN_LAEs}
\end{table*}


\section{ Distribution of \lya\ equivalent widths and its evolution} \label{sec:lya_ew}

\subsection{Measurements of  \lya\ equivalent widths and scale lengths} \label{subsec:measurements_ew}

To derive \ew\ and standard deviation values for each object, 
we performed Monte Carlo simulations. 
To do so, we first generated $300$ sets of continuum fluxes at $1216$ \AA\ and \flya\ 
based on the assumption that the distributions 
are Gaussian with mean and standard deviation values 
derived in \S \ref{sec:uv_prop_method} and \S \ref{subsec:measurement_llya}, respectively. 
We then obtained $300$ sets of \ew\ as follows: 
\begin{eqnarray}
\label{eq:rest_ew}
EW_{\rm 0} = \frac{F_{\rm Ly\alpha}}{f_{\rm \lambda, cont}} \times \frac{1}{(1+z_{\rm sp})}. 
\end{eqnarray}
For each object, the mean and standard deviation of the distribution of measurements 
are adopted as the measured and error values, respectively. 
In Table \ref{tab:quantities}, 
we list the mean, median, standard deviation, and standard error values of \ew\ for our entire sample.

\begin{figure*}[tbp]
\centering
\includegraphics[width=21cm]{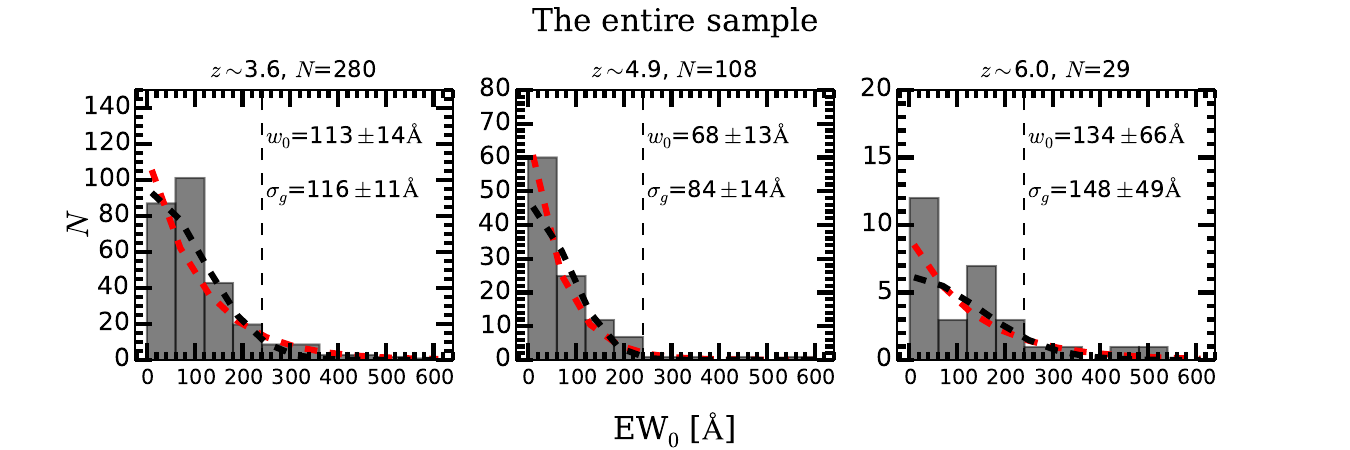}
\caption[]
{
From left to right, gray histograms correspond to 
\ew\ distributions for $z\sim3.6$, $4.9$, and $6.0$ 
with a bin width of $60$ \AA. 
One (one) object at $z\sim3.6$ ($4.9$) with \ew\ $> 600$ \AA\ 
is placed at \ew\ $= 600$ \AA\ for display purposes. 
The vertical dashed line indicates \ew $= 240$ \AA\  
(cf. \citealt{schaerer2003, raiter2010}).
The red dashed lines show the best-fit curves of the distributions expressed as 
$N=N_{\rm 0}$ exp($-$\ew/$w_{\rm 0}$), 
where $w_{\rm 0}$ indicates the best-fit scale factor. 
The black dashed lines indicate the best-fit curves of the distributions expressed as 
$N= N_{\rm 0}$ exp($-$\ew$^{2}$/2$\sigma_{\rm g}^{2}$), 
where $\sigma_{\rm g}$ indicates the best-fit distribution width.  
}
\label{fig:ew_comb_distribution}
\end{figure*}

Figure \ref{fig:ew_comb_distribution} shows the \ew\ distribution for our LAEs. 
It is known that the \ew\ distribution can be described 
either with an exponential law, 
$N=N_{\rm 0}$ exp($-$\ew/$w_{\rm 0}$) 
(\citealt{gronwall2007, nilsson2009, guaita2010, zhenya_zheng2014}), 
or with a Gaussian law, 
$N=N_{\rm 0}$ exp($-$\ew$^{2}$/2$\sigma_{\rm g}^{2}$) (\citealt{ouchi2008, guaita2010}), 
where $w_{\rm 0}$ and $\sigma_{\rm g}$ are the scale factor and distribution width, respectively. 
For convenience, we refer to $w_{\rm 0}$ and $\sigma_{\rm g}$ as the scale lengths. 

We fitted the distributions with the exponential and Gaussian laws. 
To fit the data, we take Poisson errors into account. 
The best-fit $w_{\rm 0}$ ($\sigma_{\rm g}$) values are 
$w_{\rm 0} = 113\pm14$ ($\sigma_{\rm g} = 116\pm11$), 
$68\pm13$ ($84\pm14$), 
and $134\pm66$ \AA\ ($148\pm49$ \AA) 
for $z\sim3.6$, $4.9$, and $6.0$, respectively
\footnote
{
It is not trivial to determine the appropriate number of histogram bins. 
We applied various bin numbers ranging from $6$ to $15$. 
The results are well consistent with each other within uncertainties. 
The bin number in Figure \ref{fig:ew_comb_distribution} is $10$. 
}.

\subsection{Selection cut effects on the distribution of  \lya\ equivalent widths} \label{subsec:cuts_effects}

Before comparing our scale lengths ($w_{\rm 0}$ and $\sigma_{\rm g}$) 
with those in previous studies, 
we investigated how the values can be affected by 
the selection of LAEs (i.e., limiting UV magnitudes, \lya\ luminosities, and \ew). 
Indeed, previous studies have shown that fainter \muv\ objects have larger \ew\  (e.g., \citealt{ando2006, ouchi2008} 
, see also \S \ref{sec:ando_effect}) 
and that there might be a correlation between \llya\ and \ew\ 
(Figure $9$ of \citealt{gronwall2007}).
Thus,  the scale lengths may change with different selection cuts, 
as pointed out by \cite{garel2015}. 
Because our LAEs span wide ranges of \muv and \llya, 
we were able to study all of these effects. 

\begin{figure*}[tbp]
\centering
\includegraphics[width=20cm]{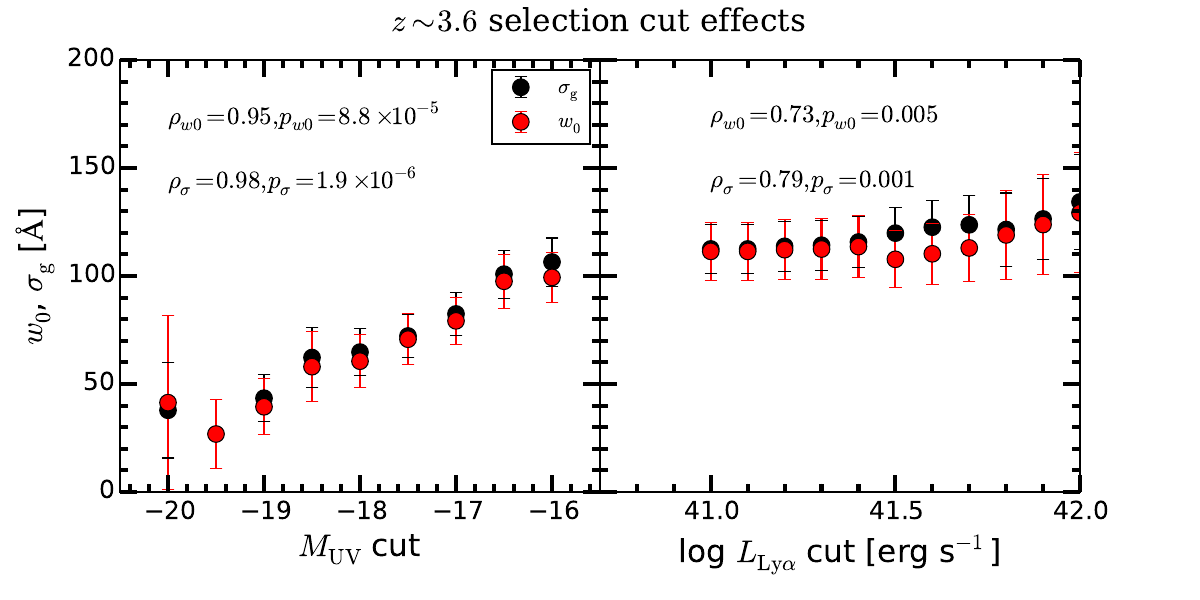}
\caption[]
{
From left to right, $w_{\rm 0}$ and $\sigma_{\rm g}$ 
for various cuts in \muv\ and log \llya\ at $z\sim3.6$. 
The red and black circles represent 
the scale factor ($w_{\rm 0}$) and distribution width ($\sigma_{\rm g}$), respectively. 
In each panel, $\rho_{w0}$ ($\rho_{\sigma}$) indicates 
the Spearman rank correlation coefficient for the relation 
in the case of the exponential (Gaussian) law, 
while $p_{w0}$ ($p_{\sigma}$) denotes the probability satisfying the null hypothesis. 
}
\label{fig:scale_lengths_different_cuts}
\end{figure*}

To do so, we remeasured \ew\ scale lengths of our LAEs with various selection cuts. 
As an example, Figure \ref{fig:scale_lengths_different_cuts} shows 
\ew\ scale lengths plotted against various cuts in \muv\ and \llya\ at $z\sim3.6$. 
The left panel of Figure \ref{fig:scale_lengths_different_cuts}
shows the \ew\ scale lengths for objects 
satisfying \muv\ $< M_{\rm UV}$ cut: 
i.e., we include \muv\ fainter objects as the $M_{\rm UV}$ cut value increases. 
We carried out the Spearman rank coefficient test 
to evaluate the significance of a correlation. 
In the case of the exponential (Gaussian) law, 
the rank correlation coefficient is $\rho_{w0} = 0.95$ ($\rho_{\sigma} = 0.98$), 
while the probability satisfying the null hypothesis is  $p_{w0} = 8.8\times10^{-5}$ 
($p_{\sigma} = 1.9\times10^{-6}$). 
Thus, we quantitatively show that including fainter \muv\ objects 
increases $w_{\rm 0}$ and $\sigma_{\rm g}$. 
A similar relation between \ew\ scale lengths and \muv\ cuts has been 
recently demonstrated by \cite{oyarzun2017} based on a Baysian approach. 
The right panel of Figure \ref{fig:scale_lengths_different_cuts}
shows the \ew\ scale lengths for objects 
satisfying  log \llya\ cut < log \llya: 
i.e., we include \llya\ faint objects as the log \llya\ cut value decreases. 
In the case of the exponential (Gaussian) law, 
the rank correlation coefficient is $\rho_{w0} = 0.73$ ($\rho_{\sigma} = 0.79$), 
while the probability satisfying the null hypothesis is  $p_{w0} = 0.005$ 
($p_{\sigma} = 0.001$). 
Although the significance level is weaker than that in the left panel, 
there is a trend that including fainter \llya\ objects decreases the scale lengths. 
The correlation is due to the fact that brighter \llya\ objects 
have larger \ew\ values at a given \muv. 
For redshift bins at $z \sim 4.9$ and $6.0$, 
we confirmed similar trends between scale lengths and selection cuts. 

We now compare the \ew\ scale factor of our LAEs, $w_{\rm 0}$, 
with those in previous studies
(\citealt{gronwall2007, ciardullo2012, zhenya_zheng2014, kashikawa2011}). 
For fair comparisons, we applied similar selection cuts as adopted in the previous studies to our LAEs, 
which are summarized in Table \ref {tab:comp_previous}. 
We take the low $z$ case as an example. 
While the $w_{\rm 0}$ value for the entire $z\sim3.6$ sample is $113\pm14$ \AA\ (Figure \ref{fig:ew_comb_distribution}), 
the $w_{\rm 0}$ value significantly reduces to $74\pm19$ \AA\ 
if we adopt the same selection cut as in \cite{gronwall2007}. 
The latter value is very consistent with that reported in \cite{gronwall2007}. 
From this table, we find that our $w_{\rm 0}$ values are consistent with those in previous studies 
within $1\sigma$ uncertainties, although these uncertainties are large at $z\sim4.9$ and $6.0$. 
The results again demonstrate that \ew\ scale lengths are sensitive to the 
selection functions of LAEs. 
The results also imply that care must be taken 
when comparing data points based on different selections.

\begin{table*}
\centering
\caption{Comparisons of $w_{\rm 0}$ in this study with those in previous studies with the same selection functions.} 
\begin{tabular}{cccccc}
\hline
Study & redshift & \muv\ Limit & log \llya\ Limit & Reference $w_{\rm 0}$ & 
$w_{\rm 0}$ in this study\\ 
& & [AB mag.] & [erg s$^{-1}$] & [\AA] & [\AA] \\
(1) & (2) & (3) & (4) &  (5) & (6)  \\
\hline
$z\sim3.6$ & &  & & \\
\hline
\cite{gronwall2007} & $3.1$ & $-18.0$ & $42.0$ & $75\pm6$ & $74\pm19$ \\ 
\cite{ciardullo2012} & $3.1$ & $-18.6$ & $42.0$ & $64\pm9$ & $60\pm20$ \\ 
\hline     
$z\sim4.9$ & & & & \\
\hline
\cite{zhenya_zheng2014} & $4.5$ & $-17.0$ & $42.4$ & $50\pm11^{a}$ ($167^{+44}_{-19}$) 
& $143\pm64$ \\
\\
\hline     
$z\sim6.0$ & & & & \\
\hline
\cite{kashikawa2011} & $5.7$ & $-18.0$ & $42.0$ & $108\pm20^{b}$ & $157\pm110$ \\
\cite{kashikawa2011} & $6.6$ & $-18.0$ & $42.0$ & $79\pm19^{b}$ & $157\pm110$ \\
\hline
\end{tabular}
\tablefoot{
Comparisons of our $w_{\rm 0}$ with those in previous studies. 
For fair comparisons, we apply to our LAEs similar selection cuts adopted in previous studies. 
\\
(1) Reference study; 
(2) typical redshift in the reference study; 
(3) lower limit of \muv\ in the reference study; 
(4) lower limit of \llya\ in the reference study; 
(5) $w_{\rm 0}$ values in the reference study; 
and 
(6) $w_{\rm 0}$ values in our LAEs with similar selection cuts of (3) and (4). 
\\
$^{a}$  The value without parentheses is  the scale factor obtained from a direct fitting to the distribution, 
while the value with parentheses indicates $w_{\rm 0}$ derived from simulations in \cite{zhenya_zheng2014}. 
\\
$^{b}$ Since the scale factors are not listed in \cite{kashikawa2011},
we take the values from \cite{zhenya_zheng2014} who fitted the 
\ew\ distribution of LAEs in \cite{kashikawa2011}. 
}
\label{tab:comp_previous}
\end{table*}


\subsection{Evolution of \ew\ scale lengths} \label{subsec:evolution_ew}

We examined the redshift evolution of the \ew\ scale lengths. 
For fair comparisons of the scale lengths at different redshifts, 
we need to take into account the fact that 
lower $z$ data are deeper than high $z$ data 
and that the \udft\ field is deeper than the \mosaic\ field in \lya. 
To take these into account, we only included LAEs with \muv\ $< -18.5$
and log \llyaunil\ $>42.2$
(see black dashed lines in Figure \ref{fig:MUV_LLyA_literature}). 
In these ranges, we are left with $40$, $31$, and $16$LAEs 
at $z\sim3.6$, $4.9$, and $6.0$, respectively. 
We obtain scale factors of 
$w_{\rm 0} = 71\pm19$, $81\pm36$, and $107\pm94$ \AA\ 
at $z\sim3.6$, $4.9$, and $6.0$, respectively. 
Likewise, we obtain distribution widths of 
$\sigma_{\rm g} = 73\pm19$, $87\pm28$, and $148\pm93$ \AA\ 
at $z\sim3.6$, $4.9$, and $6.0$, respectively.

\begin{figure*}[tbp]
\centering
\includegraphics[width=14cm]{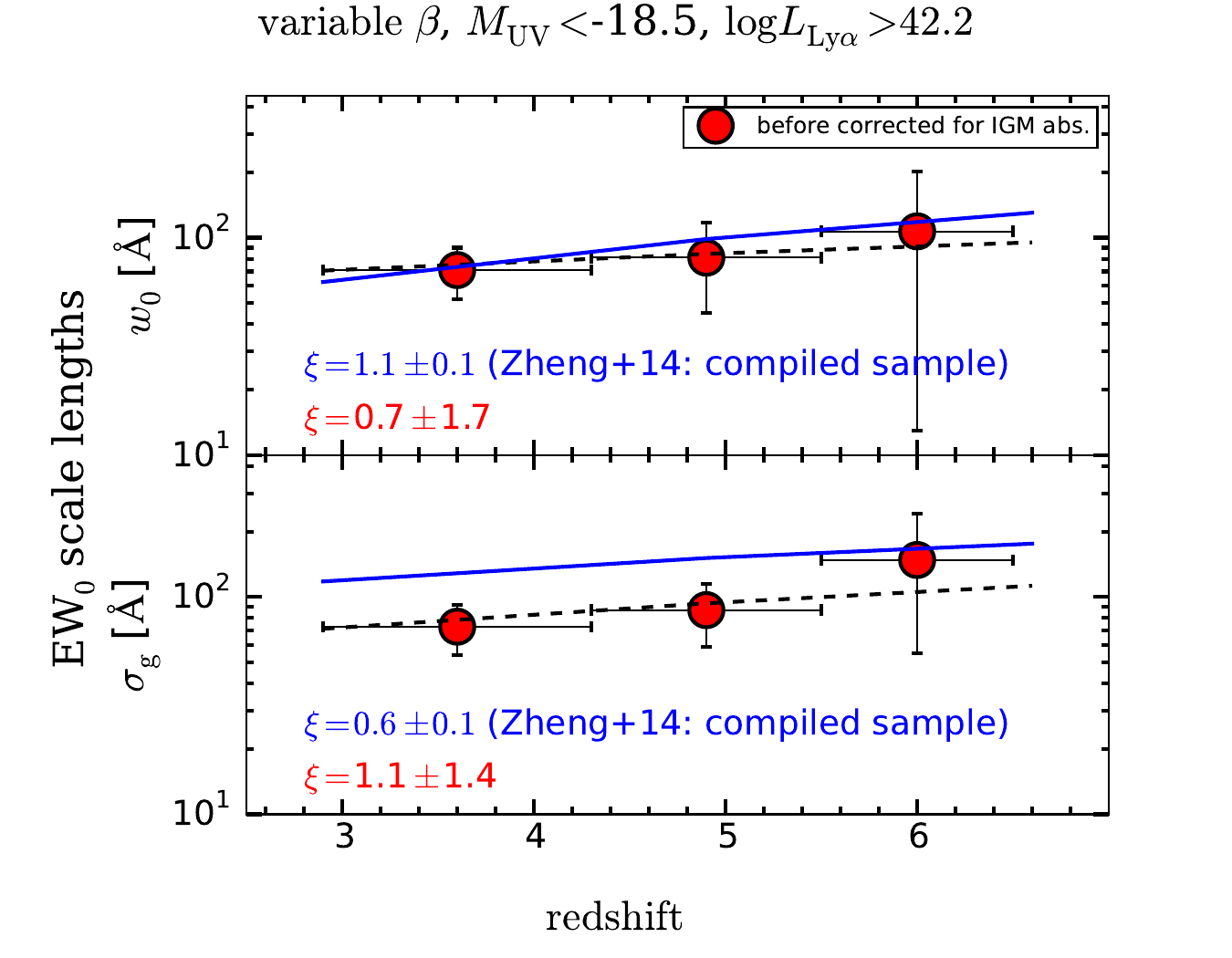}
\includegraphics[width=14cm]{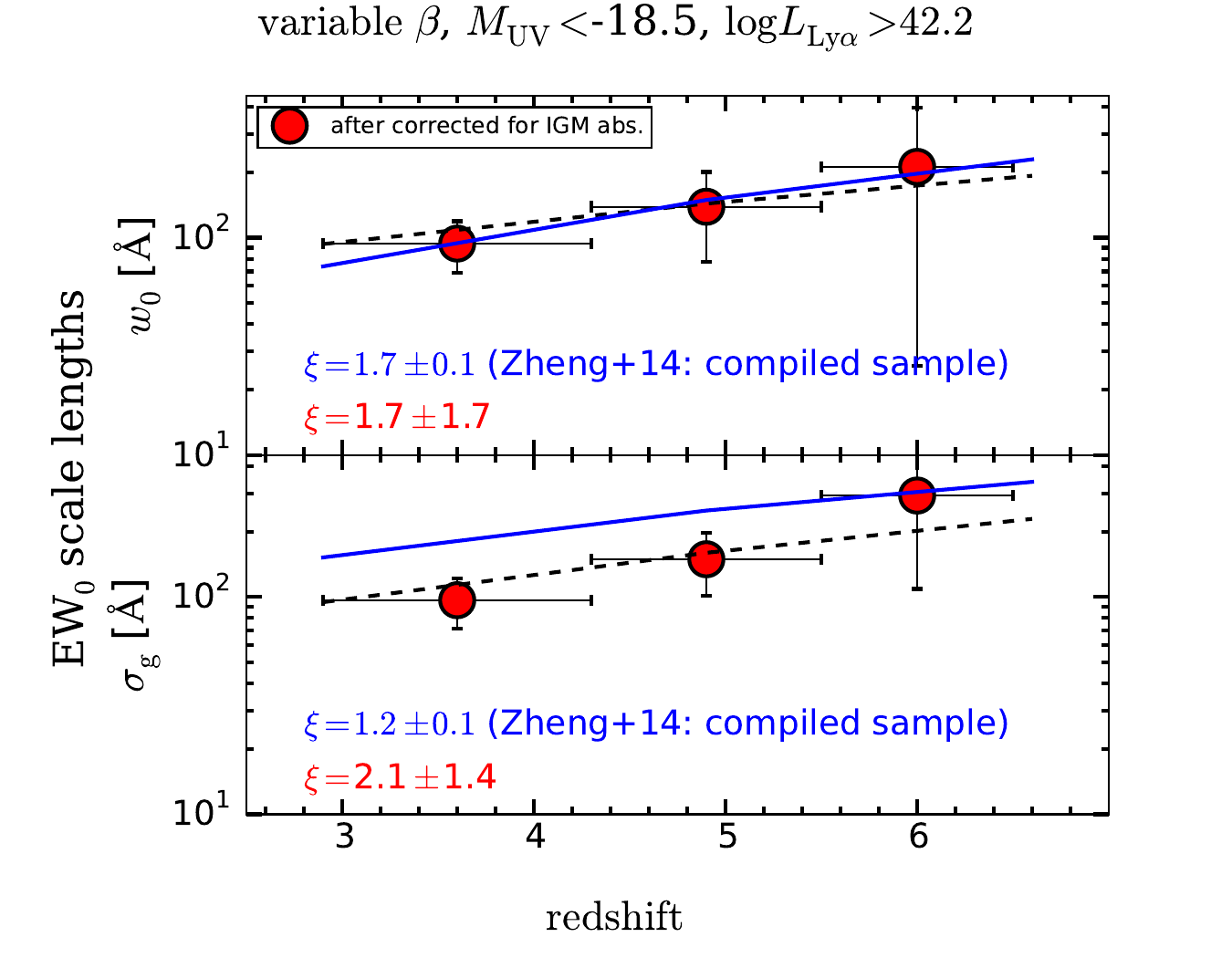}
\caption[]
{
Top two panels: Evolution of the scale lengths 
before the IGM attenuation correction on \lya. 
In this study, only LAEs with \muv\ $<-18.5$ and log \llya\ $>42.2$  
are used for fair comparisons at $2.9 < z < 6.6$ 
(see dashed lines in Figure \ref{fig:MUV_LLyA_literature}). 
The black dashed curves show the best fit to our data points 
expressed as $A\times(1+z)^{\xi}$, 
while the blue curves shows the best fit obtained in \cite{zhenya_zheng2014} 
with a compiled sample of LAEs at $0 < z < 7$ that has different selection functions. The 
$\xi$ value indicates the significance of the redshift evolution of the scale lengths. 
The bottom two panels indicate the evolution of the scale lengths 
after the IGM attenuation correction on \lya. 
Prescriptions of \cite{inoue2014} were used 
for the IGM attenuation correction on \lya. 
The meanings of the curves are the same as those in the top panels. 
}
\label{fig:evolution_scale_lengths}
\end{figure*}

In the top two panels of Figure \ref{fig:evolution_scale_lengths}, 
we plot the redshift evolution of the scale lengths of our LAEs. 
The red circles show the redshift evolution for the objects 
with \muv\ $<-18.5$ and \llyaunil\ $>42.2$. 
These scale lengths are apparent values. 
To correct for IGM attenuation at wavelengths shorter than $1215.67$ \AA, 
we used the prescriptions of \cite{inoue2014}, 
which are updated versions of those of \cite{madau1995}. 
At $z\sim3.6$, $4.9$, and $6.0$, \lya\ transmission at wavelengths 
shorter than $1215.67$ \AA\ is $0.51$, $0.17$, and $0.01$, respectively.  
Correcting  our apparent scale lengths with these factors, 
we obtain intrinsic $w_{\rm 0}$ ($\sigma_{\rm g}$) values of 
$94\pm25$ ($97\pm25$), $139\pm62$ ($149\pm48$), and $212\pm186$ ($293\pm184$) \AA\ 
at $z\sim3.6$, $4.9$, and $6.0$, respectively. 
In the bottom two panels of Figure \ref{fig:evolution_scale_lengths}, 
the red circles indicate the redshift evolution of the scale lengths 
corrected for the IGM attenuation on \lya.

Following \cite{zhenya_zheng2014}, we evaluated the redshift evolution of the scale lengths 
in the form of $w_{\rm 0}, \sigma_{\rm g} = A \times (1+z)^{\xi}$, 
where ${\xi}$ values indicate the strength of the redshift evolution. 
In the top two panels, before the IGM correction, 
we obtain the $\xi$ value of $w_{\rm 0}$ ($\sigma_{\rm g}$) to be $0.7\pm1.7$ ($1.1\pm1.4$). 
In the bottom two panels, after the IGM correction, 
we obtain the $\xi$ value of $w_{\rm 0}$ ($\sigma_{\rm g}$) to be $1.7\pm1.7$ ($2.1\pm1.4$). 
The best-fit curves are shown as black dashed lines in Figure \ref{fig:evolution_scale_lengths}. 
Owing to the large error bars in the $\xi$ values, 
we cannot conclude if the redshift evolution of the scale lengths exists. 
Our $\xi$ values are consistent with the values presented by \cite{zhenya_zheng2014} 
within $1\sigma$ uncertainties. 
\cite{zhenya_zheng2014} claimed a strong redshift evolution of scale lengths at $0 < z < 7$ 
based on a compiled sample of their LAEs and those from the literature. 
The authors obtained $\xi$ values of $w_{\rm 0}$ to be $1.1\pm0.1$ ($1.7\pm0.1$) 
before (after) IGM correction. 
The small uncertainties in $\xi$ values in  \cite{zhenya_zheng2014} 
are due to the large number of data points taken from the literature. 
However, we caution that 
the compiled sample of \cite{zhenya_zheng2014} have complicated selection cuts; 
the different data points from the literature have different selection cuts. 
For example, the literature with different selection cuts listed in Table \ref{tab:comp_previous} 
are included in these studies. 
Therefore, although our $\xi$ values are consistent with those of \cite{zhenya_zheng2014} 
we need a large number of LAEs with a uniform selection function 
at $0<z<7$ for a definitive conclusion (see also \citealt{shibuya2017}).

There are two assumptions in the correction of IGM attenuation on 
\lya,\ as discussed in \cite{ouchi2008}. 
First, the IGM attenuation prescription that we used (\citealt{inoue2014}) 
computes the \textit{mean} \lya\ transmission at a given redshift. 
Observations of $z\sim2-3$ dropouts show that 
HI absorption is enhanced near galaxies owing to their biased locations (\citealt{rakic2012, turner2014}). 
If the same trend is also true for our LAEs, we may underestimate the effect of IGM attenuation. 
In this scenario, the true redshift evolution of the intrinsic scale lengths 
might be stronger than the evolution 
we show in bottom two panels of Figure \ref{fig:evolution_scale_lengths}. 
Second, we assumed that the intrinsic \lya\ profiles are symmetric around the line center 
and we applied the IGM attenuation factor to the blue side of the \lya\ line only. 
However, it is well known that the peak of the \lya\ line is often redshifted 
with respect to the systemic redshift 
(e.g., \citealt{steidel2010, rakic2011, hashimoto2015, henry2015, inoue2016, stark2017}, 
also Verhamme et al. 2017 in prep.), 
which is often interpreted as a signature \lya\ transfer effects in galactic winds. 
Theoretical studies have shown that the impact of IGM attenuation can be significantly reduced 
in the case where the \lya\ line emerging from galaxies is redshifted 
by a few hundreds of km s$^{-1}$ (\citealt{haiman2002, dijkstra2011, choudhury2015, garel2012, garel2016}). 
Interestingly,  \cite{hashimoto2013, shibuya2014b, erb2014}  showed that the \lya\ velocity offset 
is smaller for larger \ew\ objects.
Therefore, the true IGM attenuation correction would be larger for larger \ew\ objects. 
In this case, the true evolution of intrinsic \ew\ scale lengths might be stronger 
than the evolution we show in bottom two panels of Figure \ref{fig:evolution_scale_lengths}. 

To summarize, 
our data points alone cannot conclude if redshift evolution of 
the observed \ew\ scale lengths exists. 
However, IGM correction effects are likely to strengthen 
the redshift evolution in intrinsic \ew\ scale lengths. 
We again stress that it is important to take selection function effects into account.

\subsection{Assumption of flat $\beta$ in estimates of \ew} \label{subsec:beta_assumption}

Many previous studies have assumed flat UV continuum slopes ($\beta=-2.0$) 
to derive continuum fluxes at $1216$ \AA\ 
(e.g., \citealt{malhotra2002, shimasaku2006, guaita2011, mawatari2012, zhenya_zheng2014, shibuya2017}). 
We examine how this assumption affects the redshift evolution of scale lengths. 
As shown in Table \ref{tab:quantities}, 
the typical $\beta$ value is shallower than $-2.0$ at $z\sim3.6$. 
Thus, if we assume a flat $\beta$ at $z\sim3.6$, 
the continuum fluxes at $1216$ \AA\ are overestimated, 
which in turn leads to underestimates of \ew. 
In contrast, at $z\sim4.9$ and $6.0$, 
typical $\beta$ values are steeper than $-2.0$ 
and consequently the \ew\ values are overestimated.
These effects therefore naturally lead to underestimates (overestimates) 
of the scale lengths at $z\sim3.6$ ($z\sim4.9$ and $6.0$). 
It is then possible that this can strengthen the redshift evolution of the scale lengths. 

To evaluate this, we re-examined 
the strength of the redshift evolution, $\xi$, under the assumption of $\beta=-2.0$. 
Because of the large error bars in $\xi$ values, 
the results are consistent with those with variable $\beta$. 
Table \ref{tab:flat_assumption} summarizes the \ew\ statistics 
and scale lengths for the two cases of variable and fixed $\beta$. 
Although our limited sample does not show the significant impact of 
the flat $\beta$ assumption, future works would need to consider variable $\beta$
to remove possible systematics.

\begin{table*}
\centering
\caption{Summary of the influence of a variable/flat $\beta$ slope.}
\begin{tabular}{cccccc}
\hline
Method & mean \ew & median \ew & $\sigma$ & $w_{\rm 0}$ (all)& $w_{\rm 0}$ (\muv\ $<-18.5$)\\ 
(1) & (2) & (3) & (4) &  (5) & (6) \\
\hline
$z\sim3.6$ & &  & &\\
\hline
variable $\beta$ & $113$ & $87$ & $96$ 
& $113\pm14$ 
& $63\pm15$ 
\\ 
flat $\beta$ & $96$ & $79$ & $79$ 
& $92\pm11$ 
& $50\pm11$  
\\ 
\hline 
$z\sim4.9$ & &  & &\\
\hline 
variable $\beta$ & $83$ & $56$ & $88$ 
& $68\pm13$ 
& $68\pm27$ 
\\ 
flat $\beta$ & $82$ & $60$ & $60$ 
& $80\pm17$ 
& $60\pm20$  
\\ 
\hline 
$z\sim6.0$ & &  & &\\
\hline 
variable $\beta$ & $130$ & $97$ & $120$ 
& $134\pm66$ 
& $88\pm78$ 
\\ 
flat $\beta$ & $128$ & $76$ & $106$ 
& $130\pm67$ 
& $87\pm51$  
\\ 
\hline 
\end{tabular}
\tablefoot{
Comparisons of \ew-related values in two methods: 
one based on realistic variable $\beta$ and the other based on the assumption of $\beta=-2.0$. 
\\
(1) Method; 
(2) (3) (4) mean, median, and standard deviation of \ew; 
(5) scale factors for the entire sample; 
and 
(6) scale factors for the limited subsample with \muv\ $<-18.5$. 
}
\label{tab:flat_assumption}
\end{table*}


\section{Ando effect} \label{sec:ando_effect}

\begin{figure*}[tbp]
\centering
\includegraphics[width=20cm]{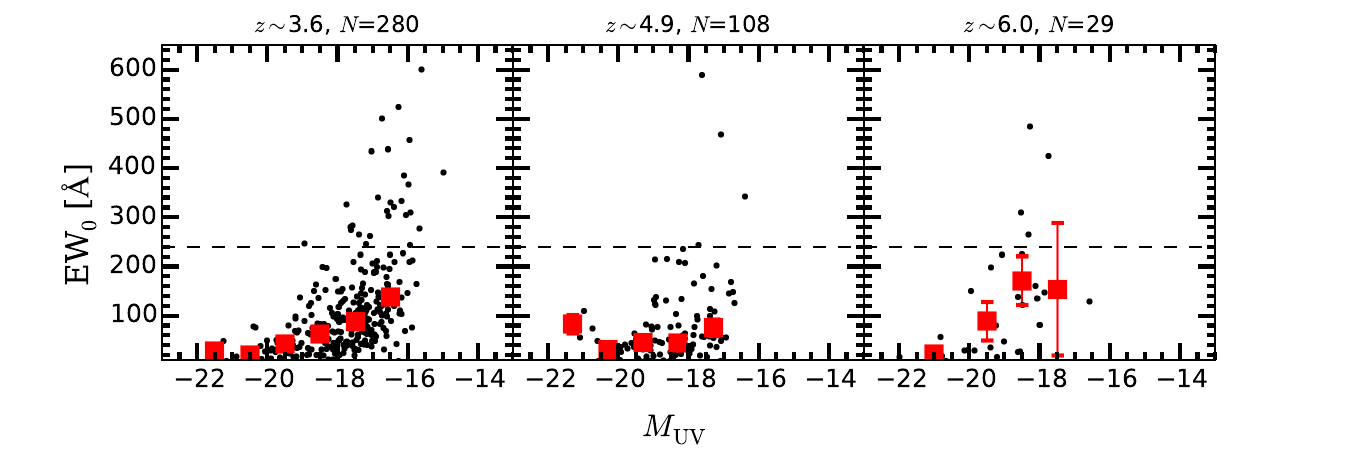}
\caption[]
{
From left to right, \ew\ plotted against \muv\ 
for $z\sim3.6$, $4.9$, and $6.0$. 
The black circles indicate our individual LAEs.
One (one) objects at $z\sim3.6$ ($4.9$) with \ew\ $> 600$ \AA\ 
are placed at \ew\ $= 600$ \AA\ for display purposes. 
The horizontal dashed line indicates \ew\ $= 240$ \AA\  
(cf. \citealt{schaerer2003, raiter2010}).
The red squares show the biweight mean values  
of \ew\ in each \muv\ bin.
}
\label{fig:MUV_EW}
\end{figure*}

We now turn our attention to the relation between \ew\ and \muv. 
As can be seen from Figure \ref{fig:MUV_EW}, 
bright continuum objects are always associated with low \ew\ values 
while UV-faint galaxies span a wide range of \ew, and some of these galaxies turn out to be very strong emitters. 
The biweight mean and error values for each magnitude bin are listed in Table \ref{tab:biweight_sample}.  
This trend was found by \cite{ando2006} for LBGs at $z\sim5-6$ 
and is confirmed by later studies of high $z$ LAEs and LBGs at $z\sim3-7$ 
(LAEs: e.g., \citealt{shimasaku2006, ouchi2008, furusawa2016, ota2017}; 
LBGs: e.g., \citealt{stark2010}). 
Following the previous studies, we refer to this effect as the ``Ando effect''. 

While several physical reasons have been invoked to interpret this trend 
(e.g., \citealt{garel2012, verhamme2012}), 
some studies have argued that it can be completely attributed to selection effects. 
\cite{nilsson2009} argued that 
the lack of small \ew\ at faint \muv\ is due to limiting \lya\ values, 
whereas the lack of large \ew\ at bright \muv\ is caused by their rarity, i.e., small survey areas 
(see also \citealt{jiang2013, zhenya_zheng2014}).

\begin{figure}[tbp]
\centering
\includegraphics[width=8cm]{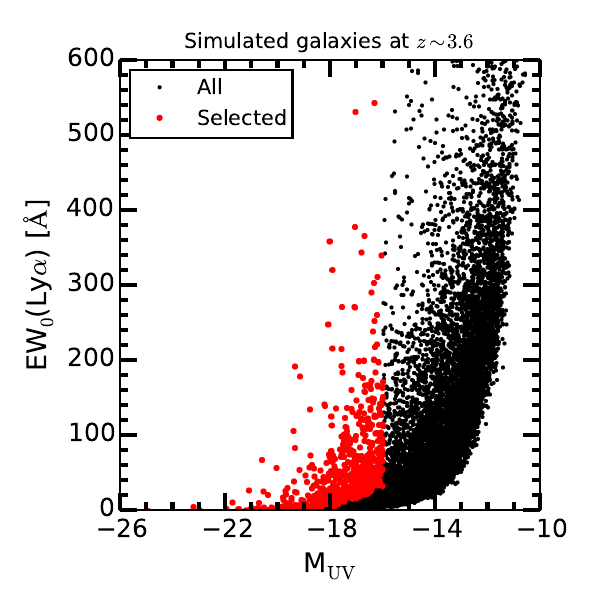}
\caption[]
{
\lya\ equivalent widths plotted against \muv\ for simulated objects $z\sim3.6$. 
The black dots show all mock galaxies with a limiting \lya\ luminosity of log \llya\ $=40.0$. 
The red circles indicate mock galaxies after the selection cuts 
of \muv $<-16.0$ and log \llya $> 41.0$ to mimic our observations. 
}
\label{fig:MC_MUV_EW}
\end{figure}

We examined whether our selection technique generates the Ando effect 
based on Monte Carlo simulations. 
We take an example of the result in the low-$z$ bin, $2.90 < z < 4.44$. 
First, we generated random log \llya\ values that follow the observed \lya\ luminosity function in D17. 
The \lya\ luminosity range is set from log \llya\ $=40.0$ to $44.0$ erg s$^{-1}$ 
with a bin size of log \llya\ $=0.1$ erg s$^{-1}$. 
Based on the results of D17, we assumed 
log $L^{*} = 42.59$ erg s$^{-1}$, log $\phi^{*} = -2.67$ Mpc$^{-3}$, and $\alpha = -1.93$, 
where $L^{*}$, $\phi^{*}$, and $\alpha$ represent the characteristic luminosity, 
characteristic amplitude, and slope of the Schechter function, respectively. 
Second, we generated random \ew\ values that follow the exponential distribution. 
We assumed a scale length of  $w_{\rm 0} = 113$ \AA\ 
based on our results at $2.90 < z < 4.44$ (Figure \ref{fig:ew_comb_distribution}). 
Third, we generated random $\beta$ values that follow a Gaussian distribution 
with mean and standard deviation values in Table \ref{tab:quantities}. 
Finally, redshift values are drawn from the uniform random distribution between $2.90 < z < 4.44$. 
On the assumption that \llya, \ew, $z$, and $\beta$ do not correlate with each other, 
we assigned these numbers to each $10,000$ mock galaxy. 
We estimated \muv\ values in the opposite way as equations (\ref{eq:beta}) to (\ref{eq:rest_ew}). 
In Figure \ref{fig:MC_MUV_EW}, 
the black dots show all $10,000$ simulated galaxies. 
To mimic our observations, we imposed selection cuts of log \llya\ $> 41.0$ and \muv\ $<-16.0$ 
on the mock galaxies based on the left panel of Figure \ref{fig:MUV_LLyA_literature}. 
The selected objects are denoted as red circles. 
As can be seen from Figure \ref{fig:MC_MUV_EW}, 
the lower boundary of the relation is created due to the limiting \llya\ value. 
On the other hand, the upper boundary of the relation is due to the rarity of \muv\ bright 
objects with large \ew\ values. 
These results are consistent with those in, for example, \cite{zhenya_zheng2014}. 
Based on these results, we conclude that 
we cannot rule out the possibility 
that the Ando effect is completely due to the selection bias 
if our assumptions are correct.

\section{Very large \lya\ equivalent width\ \lya\ emitters} \label{sec:large_ew}

\subsection{$12$ Very large \lya\ equivalent width\ \lya\ emitters} \label{subsec:large_lae}

Very large \ew\ LAEs are interesting because 
they are candidates for galaxies 
in the early stages of the galaxy formation and evolution (\citealt{hashimoto2017} and references therein). 
In this study, we define very large \ew\ LAEs as objects having \ew\ $> 200$ \AA. 
We list $12$ LAEs with \ew\ $> 200$ \AA\ above $1\sigma$ uncertainties 
in Table \ref{tab:large_ew_LAEs}. 
To investigate the significance, 
we calculated $\sigma_{\rm 200}=$ (\ew $- 200$)/\ew$_{\rm err.}$, 
where \ew$_{\rm err.}$ is the error value of \ew. 
The values range from $\sigma_{\rm 200} = 1.0$ to $2.8$. 

We compared our $\sigma_{\rm 200}$ 
with those in previous studies that focus on properties of very large \ew\ LAEs. 
\cite{kashikawa2012} reported a spectroscopically identified 
very large \ew\ LAE at $z=6.5$. 
The object has \ew\  $= 436^{+422}_{-149}$ \AA
\footnote{The value given in \cite{kashikawa2012}, EW $=872^{+844}_{-298}$ \AA, 
is after the correction for the IGM attenuation on \lya. 
For a fair comparison with our values, 
we used the \ew\ before the correction for IGM attenuation. 
}, 
corresponding to $\sigma_{\rm 200}=1.6$. 
\cite{sobral2015} reported a very bright LAE at $z=6.6$, CR7, 
whose \lya\ is spectroscopically identified. 
This object has \ew\ $=211\pm20$ \AA, corresponding to $\sigma_{\rm 200}=0.6$. 
Recently, \cite{hashimoto2017} have investigated six $z\sim2$ LAEs with very large \ew.
In this study, four objects have \ew\ $>200$ \AA\ 
with $\sigma_{\rm 200} = 0.7-5.3$. 
Therefore, our $\sigma_{\rm 200}$ are similar to 
those in previous studies that focus on very large \ew\ LAEs. 
Among the $12$ very large \ew\ LAEs, 
four objects, ID3475, ID4623, ID6376, and ID4598, 
have extremely large \ew\ $\gtrsim400-600$ \AA. 
In these objects, \ew\ and $\sigma_{\rm 200}$ values are comparable to 
or higher than the very large \ew\ LAE studied by \cite{kashikawa2012}. 

Also, our very large \ew\ LAEs have relatively shallow $\beta$ values, 
$-1.6\pm0.1$, where uncertainty denotes the standard error. 
The result can indicate the presence of nebular continuum (\citealt{schaerer2003}). 
Alternatively, these red $\beta$ values can also indicate that 
our LAEs are affected by hidden AGN activity. 

In the following sections, we investigate two very large \ew\ LAEs 
whose \ew\ can be explained by mergers or hidden type-II AGN activity. 
We discuss possible explanations for the remaining $10$ very large \ew\ LAEs in \S \ref{subsec:lya_sf}. 

\subsection{ID7159: Detection of C{\sc iv} $\lambda 1549$ $-$ an AGN-like\ \lya\ emitter?} \label{subsec:id7159}

Object ID7159, at $z=3.00$, 
has \muv\ $= -17.6\pm0.1$ and \ew\ $=286\pm85$ \AA.  
The object has a detection of the C{\sc iv} $\lambda 1549$ line, 
but does not have detections of the C{\sc iii}] $\lambda 1908$ nor He{\sc ii} $\lambda 1640$ lines. 
Since the C{\sc iv} line is often associated with AGN activity, 
it is possible that hidden AGN activity produces additional ionizing photons 
(e.g., \citealt{malhotra2002, dawson2004}). 
However, \cite{stark2015b} revealed that 
the C{\sc iv} line can also be emitted by a young stellar population with very hot metal-poor stars
(see also \citealt{christensen2012, mainali2017, schmidt2017}). 
Indeed, ID7159 has \lya\ FWHM of 464 km s$^{-1}$ after the instrumental correction, 
which is similar to the typical FWHM value of $z\sim2-3$ LAEs, $100-500$ km s$^{-1}$ 
(e.g., \citealt{trainor2015, hashimoto2017}). 
Therefore, it is difficult to conclude whether ID7159 is a star-forming LAE or an AGN-like LAE 
with the current data.  
\footnote{
The X-ray flux upper limits of ID7159 are 
$1.9 \times 10^{-17}$, $6.4 \times 10^{-18}$, and $2.7 \times 10^{-17}$ 
erg cm$^{-2}$ s$^{-1}$ in the three bands at 
$0.5-7.0$ keV, $0.5-2.0$ keV, and $2-7$ keV, respectively (see \S \ref{sec:AGN}). 
}

\subsection{ID7283:  Merger activity?} \label{subsec:merger}

Interestingly, we found that one of the $12$ very large \ew\ LAEs, ID 7283, 
has a companion LAE, ID 6923, at a similar redshift. 
The projected distance between the pair is $\sim 25$ kpc. 
Based on the \lya\ narrowband image, 
we confirmed that \lya\ emission of the pair LAEs are well separated 
and not contaminated by the \lya\ emission of the companion. 

It is possible that \lya\ emission in the pair LAEs is powered by collisional excitation 
followed by merger activity and subsequent gravitational cooling 
(e.g., \citealt{taniguchi2000, oti-floranes2012, rosdahl2012}). 
Alternatively, the pair might create strong ionizing fields 
that serve as external UV background sources for each object,
leading to additional fluorescent \lya\ emission from circum-galactic gas.

\begin{table*}
\centering
\caption{Properties of $12$ very large \ew\ LAEs, \ew\  $>200$ \AA.} 
\begin{tabular}{cccccccl}
\hline
ID & $z$ & \ew & $\sigma_{\rm 200}$ & log \llya & \muv  & $\beta$ & comment\\ 
(1) & (2) & (3) & (4) &  (5) & (6) & (7) & (8)\\
\hline
$489$ & $4.16$ & $362\pm86$ & $1.9$ & $42.2\pm0.1$ & $-16.8\pm0.1$ & $-1.5\pm0.3$ &  \\
$1969$ & $4.08$ & $245\pm18$ & $2.5$ & $42.9\pm0.1$ & $-18.9\pm0.1$ & $-1.9\pm0.1$ & \\
$3034$ & $4.26$ & $321\pm52$ & $2.3$ & $42.5\pm0.1$ & $-17.8\pm0.1$ & $-1.4\pm0.2$ & \\
$3475$ & $3.16$ & $495\pm104$ & $2.8$ & $42.3\pm0.1$ & $-16.7\pm0.1$ & $-1.7\pm0.2$ & \\
$4231$ & $3.47$ & $315\pm101$ & $1.1$ & $42.1\pm0.1$ & $-16.6\pm0.2$ & $-2.1\pm0.4$ & \\
$4515$ & $3.66$ & $325\pm122$ & $1.6$ & $42.0\pm0.1$ & $-16.4\pm0.1$ & $-2.1\pm0.4$ & \\
$4623$ & $3.55$ & $569\pm225$ & $1.6$ & $42.2\pm0.1$ & $-16.2\pm0.2$ & $-1.6\pm0.5$ & \\
$6376$ & $4.29$ & $441\pm149$ & $1.6$ & $42.3\pm0.1$ & $-17.0\pm0.1$ & $-1.3\pm0.6$ & \\
$7159$ & $3.00$ & $286\pm85$ & $1.0$ & $42.4\pm0.1$ & $-17.6\pm0.1$ & $-2.1\pm0.4$ & \S \ref{subsec:id7159}; weak CIV \\
$7191$ & $3.18$ & $267\pm67$ & $1.0$ & $42.3\pm0.1$ & $-17.6\pm0.1$ & $-1.1\pm0.3$ & \\
$7283$ & $3.43$ & $266\pm60$ & $1.1$ & $42.2\pm0.1$ & $-17.1\pm0.1$ & $-2.1\pm0.3$ & \S \ref{subsec:merger}; merger (pair ID$=6923$) \\
$4598$ & $5.77$ & $490\pm199$ & $1.5$ & $42.9\pm0.1$ & $-18.2\pm0.2$ & $-1.1\pm0.6$ & \\
\hline
\end{tabular}
\tablefoot{
IDs and physical quantities of $12$ very large \ew\ LAEs. \\
}
\label{tab:large_ew_LAEs}
\end{table*}


\section{Discussion} \label{sec:discussion}

\subsection{Limitations of this study} \label{subsec:limitations}

In this study, we excluded (1) $78$ objects with spatially multiple HST counterparts. 
Hereafter we refer to these objects as blended LAEs. 
The procedure is needed to construct a clean sample 
in which \ew\ values are robustly measured. 
For example, if we mistakenly allocated our MUSE \lya\ emission to an HST counterpart, 
the \ew\ values would be incorrect as well. 
In addition, we excluded (2) $176$ objects with a spatially single HST counterpart, 
but which do not have enough (to be defined) multicolor images. 
Hereafter, we refer to these objects as very UV faint LAEs. 
This procedure is also needed to derive \ew\ values 
with small systematic uncertainties introduced by the flat $\beta$ ($\beta=-2.0$) assumption.  
However, since the number fraction of theses LAEs are not negligible 
($11\%$ and $26\%$), we discuss possible bias effects 
introduce by excluding these objects (see \S \ref{subsubsec:our_sample}).

To examine the first point,  we compared \lya\ fluxes of the two samples: 
blended LAEs and non-blended LAEs. 
To do so, we performed a two-sample K-S test. 
We find that the $p-$value is $0.0001$, indicating that 
the \lya\ flux distributions of the two samples are statistically different with each other. 
Likewise, we compared HST magnitudes of the two samples based on a K-S test. 
In this analysis, we used HST magnitudes of the nearest counterpart.
We take F775W, F105W, and F125W as examples. 
We find that the $p-$values are $< 0.0001$ in these HST wave bands, 
indicating that the HST magnitude distributions of the two samples are statistically different. 
These results suggest that excluding blended LAEs 
can introduce a bias effect 
in terms of \lya\ fluxes and HST magnitudes (thus \muv). 
More specifically, we find that \lya\ fluxes and HST magnitudes 
are brighter in blended LAEs than in non-blended LAEs. 
Because we cannot allocate our MUSE \lya\ emission 
to one of the HST counterparts in these cases, 
we cannot obtain accurate \ew\ measurements. 
Under the assumption that the brightest HST counterpart is responsible 
for the MUSE \lya\ emission, we could obtain lower limits of \ew. 
We leave these analyses to future works 
and stress that possible bias effects can change our results. 
Nevertheless, we can discuss the blending effects  
because of the high spatial resolution of HST. 
For example, observations based on ground telescopes alone 
cannot easily investigate these effects due to their limited spatial resolutions. 
In this sense, these results are our current best efforts.

We also examine the second point, very faint UV LAEs. 
Because these very faint UV objects would have very large \ew\ values
(or at least very large lower limits of \ew\ values given the Ando effect), 
the actual \ew\ distributions can be different from what we show in Figure \ref{fig:ew_comb_distribution}. 
In our discussion, the redshift evolution of \ew\ scale lengths can be affected by this effect.  
However, as we described in Sec. \ref{subsec:evolution_ew}, 
as long as we use sufficiently bright objects, 
our discussion remains unchanged. 
The detailed properties of these very faint UV LAEs will presented 
in M. Maseda et al. (in preparation).

\subsection{\lya\ emission powered by star formation} \label{subsec:lya_sf}

 The \ew\ value encapsulates valuable information about galaxies because this value is the ratio of the \lya\ emission and stellar continuum. 
This value is however a complex quantity hard to interpret 
because its strength is determined by several aspects 
that cannot be disentangled easily. 
Hereafter, we discuss our results in the light of previous studies on \ew\ at high redshift 
with a particular focus on the comparison with theoretical predictions. 

In parallel to high-redshift galaxy surveys, much progress has been made over the last few years 
to reconcile observational constraints on LBGs and LAEs with theoretical predictions. 
Under the assumption that \lya\ photons result from hydrogen recombination in star-forming regions, 
the observed \lya\ and UV luminosity functions at $3 < z < 6$ can be reproduced 
by various cosmological hydrodynamical simulations and semi-analytic models, 
at least within the observational uncertainties (e.g., \citealt{dayal2008, orsi2012, garel2015}). 
However, these simulations often fail at reproducing quantitatively the global shape of the \ew\ distribution. 
Unlike the observed distributions that usually peak at a lower \ew\ limit (which depends on the LAE selection) 
and extend to $\gtrsim 200$ \AA, models often predict much narrower distributions and struggle to recover 
the high fraction of objects with moderately large \ew, $100-200$ \AA, (\citealt{dayal2008, garel2012}). 
Below, we discuss possible mechanisms to reproduce a higher fraction of moderately large \ew\ LAEs. 

It has been shown that assuming different IMFs mostly changes the peak value of the \ew\ distribution 
but does not increase its width (e.g., \citealt{garel2015}), 
unless one adopts evolving or spatially varying IMFs within galaxies (e.g., \citealt{orsi2012}). 
Nevertheless, at fixed IMF, \cite{forero-romero2013} hinted that the stochastic sampling of the IMF 
can induce fluctuations in the predicted \ew\ values for a given star formation event, 
hence broadening the \ew\ distributions (see also \citealt{mas-ribas2016}). 
Alternatively, bursty star formation may also help reconcile models and observations. 
\cite{garel2015} showed that bursty star formation can be more likely to be achieved 
if one increases the gas surface density threshold to trigger the formation of stars. 
This can in turn give rise to \lya-bright and \lya-quiescent phases. 
Then, at a given time, galaxies exhibit a wide range of \ew\ values between $0$ and $\approx$ 200 \AA,\ 
which depends on the time delay since the last starburst; these values are in better agreement with our observations.

\begin{figure*}[tbp]
\centering
\includegraphics[width=18cm]{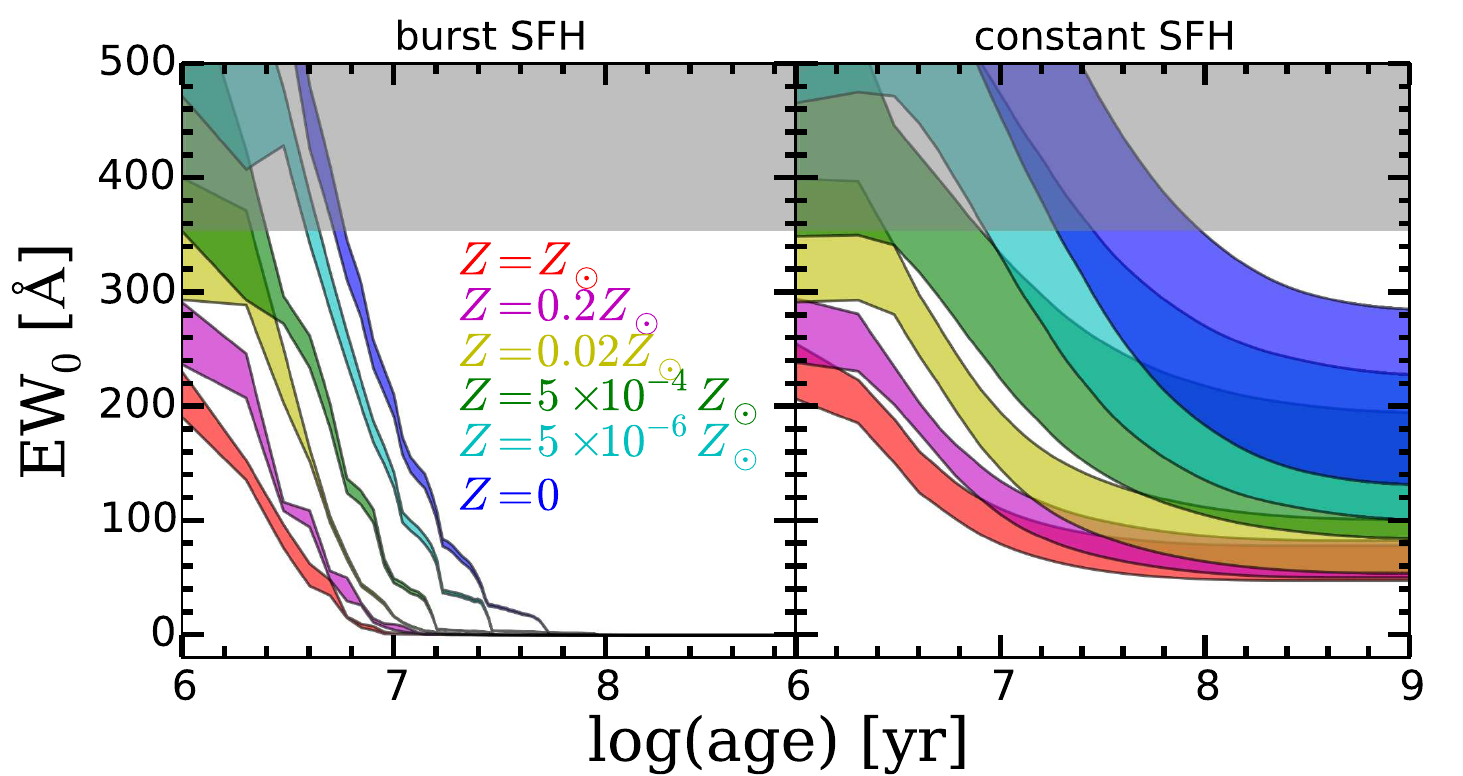}
\caption[]
{
Comparisons of the observational constraints on 
\ew\ with the models of \cite{schaerer2003, raiter2010}. 
These models show evolution of the spectral properties of 
stellar populations for stellar ages varying from $10^4$ yr to 1 Gyr 
for the starburst SFH (left panel) and the constant SFH (right panel). 
Different colors correspond to six metallicities: $Z=0$ (PopIII, blue), 
$5\times10^{-6} \ Z_{\rm \odot}$ (cyan), $5\times10^{-4} \ Z_{\rm \odot}$ (green), 
$0.02 \ Z_{\rm \odot}$ (yellow), 
$0.2 \ Z_{\rm \odot}$ (magenta), and $Z_{\rm \odot}$ (red). 
For each metallicity, the colored shaded regions denote \ew\ ranges traced by the three 
three power-law IMFs: two Salpeter IMFs ($1-100 \ M_{\rm \odot}$ and $1-500 \ M_{\rm \odot}$) 
and a Scalo IMF ($1-100 \ M_{\rm \odot}$). 
The horizontal gray shaded regions indicate the range of nine very large \ew\ LAEs
without signatures of mergers or AGNs. 
}
\label{fig:stellar_age_metal}
\end{figure*}

In addition to the problem of moderately large \ew\ LAEs discussed above, 
we demonstrated that $12$ LAEs in our sample have \ew\ values 
larger than the typical maximal value predicted by stellar synthesis models based on standard IMFs and solar metallicity 
(EW$_{\rm max} \approx 240$ \AA; see red curves in Figure \ref{fig:stellar_age_metal}). 
While two of the $12$ very large \ew\ LAEs might be AGNs or mergers (\S \ref{sec:large_ew}), 
other interpretations are required for the $10$ remaining LAEs with very large \ew\ values. 
To discuss these very large \ew\ LAEs, 
we follow the procedure in \cite{hashimoto2017} 
who have used the models of \cite{schaerer2003} and its updated version by \cite{raiter2010}
to constrain the properties of very large \ew\ LAEs. 
These models cover metallicities from Pop III to solar and a wide range of IMFs 
assuming two different star formation histories (SFH): 
an instantaneous burst (starburst SFH) and constant star formation (constant SFH). 
Given that large \ew\ LAEs have low metallicities, 
these models with fine low-metallicity grids 
are very appropriate to investigate the large \ew\ LAEs.

Figure \ref{fig:stellar_age_metal} shows the predicted \ew\ value 
as a function of age, where each curve corresponds to the \ew\ evolution for a given stellar metallicity 
and the colored shaded regions represent the range spanned by the three assumed IMFs. 
We see that higher \ew\  values are expected for younger stellar ages and lower metallicities 
for both the starburst (left panel) and the constant SFH (right panel).
In the case of a starburst, the timescale for the \lya\ line to be visible reflects the lifetime of O-type stars, 
and increases toward lower metallicities, 
reaching log(age yr$^{-1}$) $\approx 7.5$ for PopIII stars. 
For a constant SFH, the \ew\ values decrease over similar (though slightly longer) timescales 
and then settle into a nearly constant regime with the \ew\ value  ranging from $\approx$ $50$ \AA\ 
for solar metallicity to $\approx$ $300$ \AA\ for zero metallicity. 
The gray shaded regions in Figure \ref{fig:stellar_age_metal} depict the range spanned by our $10$ LAEs. 
The mean and standard error values of this subsample is 389 $\pm$ 36 \AA, 
and here, we adopt the 1$\sigma$ lower limit, $353$ \AA. 
The comparison with the model predictions shows that our very large \ew\ LAEs 
can be explained by a recent burst of star formation ($\approx$ 10 Myr) with $Z \lesssim 0.02$ $Z\odot$, 
or by a stellar population younger than $\approx$ 100 Myr (also with $Z \lesssim 0.02$ $Z\odot$) 
for a constant SFH
\footnote
{
\cite{hashimoto2017} have also used $\beta$ values to place constraints 
on stellar ages and metallicities. 
To use $\beta$, we need to correct these values with dust extinction effects. 
Currently, dust extinction values, $E(B-V)$, are not available for our LAEs. 
Thus, we assume that that our intrinsic $\beta$ values are bluer 
than those in Table \ref{tab:large_ew_LAEs}. 
We find that our relatively shallow mean $\beta$ value, $-1.6\pm0.1$, 
does not tighten the ranges of the stellar age and metallicity of our very large \ew\ LAEs. 
}. 

While these quantities are hard to constrain observationally, 
predictions from hydrodynamical simulations suggest that galaxies can exhibit lower stellar metallicities 
at higher redshift (\citealt{ma2016, taylor2016}).
In addition, our largest \ew\ values correspond to faint galaxies (\muv\ $\gtrsim -18$), 
which plausibly consist of low-mass objects. 
According to simulations, less massive galaxies tend to have more bursty SFH 
and lower stellar metallicity at a given redshift (\citealt{ma2016, sparre2017}).
Interestingly, \cite{sparre2017} show that low-mass galaxies ($\lesssim 10^9 M_{\rm \odot}$) form most their stars 
during intense bursts of star formation, whereas the time fraction spent in burst cycles (i.e., the \textit{duty cycle})
is about $10-20\%$ at all masses. 
This \textit{duty cycle} can be compared with the fraction of very strong emitters (\ew\ $\geq 200$ \AA) 
among faint galaxies in our sample ($\approx 250$ galaxies with M$_{\rm UV} \gtrsim -18$): 
$\approx 5\%$ for objects with \ew\ uncertainties above 1$\sigma$, and $\approx 13\%$ otherwise. 
Overall, bursty star formation associated with subsolar metallicities 
seem able to account for the observed \ew\ distribution, in particular the very large \ew. 
Nonetheless, in the next sections, we investigate alternative interpretations of the very large \ew\ values 
of the nine LAEs without signatures of mergers or AGNs.

\subsection{Radiative transfer and \ew\ boost} \label{subsec:lya_rt_boost}

When propagating through inhomogeneous or multiphase media, 
\lya\ photons often take a very different path compared to non-resonant continuum photons. 
Under given conditions, the \lya\ escape fraction can then become larger than the UV continuum escape fraction, 
hence boosting the observed \ew. 
For instance, in the clumpy ISM model in which dust is locked into HI clouds (\citealt{neufeld1991, hansen_oh2006}), 
\lya\ photons scatter off the surface 
of the clouds while continuum radiation can penetrate the clouds and be absorbed by dust grains. 
This scenario has notably been shown to well recover the \ew\ distributions along with the luminosity functions, 
when brought into the cosmological context using cosmological simulations or semi-analytic models 
(\citealt{kobayashi2010, shimizu2011}). 
Some numerical \lya\ transfer experiments have claimed that a significant boost of the \textit{angular-average} \ew\  
can only be achieved under physical conditions, such as metallicity, gas density, velocity, and covering fraction, which are unlikely to be representative of the ISM at high redshift (\citealt{laursen2013, duval2014}). 
Similarly, \cite{gronke2014} investigated the angular variation of the \ew\ 
and they concluded that this quantity can be strongly enhanced along a limited number of sight lines.

Similarly, the escape of \lya\ photons is found to be highly anisotropic for non-spherical gas distribution 
(e.g., discs, bipolar winds; \citealt{verhamme2012, behrens2014a, zheng2014}) 
and varies as a function of the inclination angle. 
Even in the case where the global (i.e., angle average) \lya\ escape fraction remains 
lower than that of UV photons because of \lya\ resonant scattering, 
\lya\ photons may preferentially emerge from galaxies along low HI-opacity sight lines, 
increasing the \ew\ in these directions. 
Quantitatively speaking, these simulations predict that the \ew\ can be boosted 
up to a factor of $\approx$ 3, depending on the exact geometry, HI density, and velocity fields or the amount of dust. 
Although radiative transfer effects undoubtedly play a role in shaping the \lya\ emission properties of high-redshift galaxies, 
it remains difficult to determine at which extent these are responsible for the very large \ew\ that we observe.

\subsection{Other \lya\ production channels} \label{subsec:lya_other}

The very large \ew\ values observed in our sample may also indicate objects 
for which a significant fraction of \lya\ radiation is not produced by internal star formation. 
For example, by observing around a bright quasar, \citet{cantalupo2012} found a large sample 
of very large \ew\ LAEs for which the \lya\ emission is most likely powered by fluorescence from the quasar illumination, up to a few hundred comoving Mpc$^{3}$ around the quasar 
(see also \citealt{borisova2016, marino2017}). 
To investigate this issue, we searched for quasars in and around the UDF 
using the Veron Cetty catalog
\footnote{
https://heasarc.gsfc.nasa.gov/db-perl/W3Browse/
}.  
We used the large search radius of $10$ arcmin from the center position of the UDF. 
We find that there are no nearby QSOs within $10$ comoving Mpc from our very large \ew\ LAEs. 
The result indicates that there are no detectable active QSOs in current catalogs 
that can contribute to increasing the \ew\ value of our very large \ew\  sources with fluorescence. 
However, because of light travel effects, 
we cannot exclude the possibility 
that past QSO phases in neighboring galaxies 
within a few Mpc from our very large \ew\ LAEs could be responsible 
for the Ly$\alpha$ boosting, especially if QSO phases are short but relatively frequent 
(e.g., \citealt{cantalupo2007, cantalupo2012, trainor2013, borisova2016b, marino2017}). 
In particular, if all our very large \ew\  values are due to this effect, 
this could give us potential constraints on the AGN phase duty cycle. 
We will investigate this in detail in future work.

Likewise, it is possible that nearby AGN activity 
contributes to \lya\ fluorescence. 
For the $10$ large \ew\ LAEs without signatures of mergers or AGN activity, 
we found that none of these objects have nearby AGNs (\citealt{luo2017}) within $10$ comoving Mpc. 
Therefore, it is unlikely that AGN \lya\ fluorescence contribute to the very large \ew\ values, 
although hidden type-II AGN activity might do the job. 

Another source of \lya\ emission, independent of star formation, is gravitational cooling radiation. 
This mechanism has been invoked to explain giant Lyman-alpha blobs 
(see, e.g., \citealt{haiman2000, fardal2001, dijkstra2009}). 
There exist theoretical and numerical quantitative predictions for this process, 
although large uncertainties remain. 
These predictions suggest that 
a luminosity of \llyaunil\ $\sim 10^{42}$ erg s$^{-1}$ can be produced 
by gas falling into a dark matter (DM) halo 
with a mass of $M_h \sim 3\times 10^{11}M_\odot$  
\citep{dijkstra2009, faucher-giguere2010, rosdahl2012, yajima2012}.
From Table \ref{tab:large_ew_LAEs}, we see that this can easily account for half the flux of most of our 
very large \ew\ objects. 
Therefore, gravitational mechanism would explain an \ew\ twice as large as star formation would allow. 
If this is the case, we do need neither extremely young stellar age nor low metallicity 
to explain very large \ew\ objects. 
The two brightest objects of Table \ref{tab:large_ew_LAEs} 
(ID= 1969 and 4598) have a luminosity almost an order of magnitude larger. 
If they are in a DM halo of mass of $M_h \sim 3\times 10^{11}M_\odot$, 
cooling radiation may only boost their \ew\ by $\sim 10$ \%.  
This is not quite enough to reconcile them with the star formation limit. 
Nevertheless, the quasi-linear relation between $L_{\rm Ly\alpha}$ and $M_h$ for cooling radiation 
implies that only a moderately larger DM halo host would be able to do the job. 

\section{Summary and conclusions} \label{sec:summary}

We have presented a new large data set of $417$ LAEs 
detected with MUSE at $2.9 < z < 6.6$ in the Hubble Ultra Deep Field (UDF). 
Owing to the high sensitivity of MUSE, we detected \lya\ emission  
from log \llyaunil\ $\sim 41.0$ to $43.0$. 
For the estimates of \lya\ fluxes, we adopted the curve of growth technique 
to capture the extended emission. 
Taking into account the extended \lya\ emission is important for accurate measurements of \ew\ 
because a significant fraction of \lya\ emission 
originates from the extended component, the so-called \lya\ halo (see L17). 
In addition, with deep HST photometry data in the UDF, 
we derived UV slopes ($\beta$) and continuum fluxes of our LAEs. 
The UV absolute magnitudes range from \muv\ $\sim -16.0$ to $-21.0$ ($0.01-1.0 L^{*}_{\rm z=3}$). 
The faint-end \llya\ and \muv\ values at $z\sim3.6$ and $4.9$ 
are roughly one order of magnitude 
fainter than those in previous LAE studies based on the narrowband technique (Figure \ref{fig:MUV_LLyA_literature}). 
We derived \ew\ values 
and focused on two controversial issues: 
first, the evolution of the \ew\ distribution between $z=2.9$ and $6.6$, 
and second, the existence of very large \ew\ LAEs. 
Our main results are as follows: 

\begin{itemize}

\item 
The median $\beta$ values in our LAEs are $-1.73\pm0.04$, $-2.22\pm0.15$, and $-2.31\pm0.19$ 
at $z\sim3.6$, $4.9$, and $6.0$, respectively, where error values denote the standard errors. 
The high dynamic range of \muv\ in our LAEs 
allows us to investigate $\beta$ values in as much detail as those in dropout galaxies. 
We find a trend that $\beta$ becomes steeper at faint \muv. 
The slope d$\beta$/d\muv\ of our LAEs is in good agreement with that in dropout galaxies, 
$\approx-0.1$ (\S \ref{subsec:muv_beta} and Figures \ref{fig:MUV_beta} and \ref{fig:z_dbdM}).
We also find that $\beta$ becomes steeper at high $z$.  
At both bright (\muv\ $\approx -19.5$) and faint (\muv\ $\sim -17.5$) UV magnitude bins, 
the typical $\beta$ values decrease from $\approx -1.8$ to $-2.5$ at $z\sim3.6$ and $6.0$, respectively, 
which is consistent with results for dropout galaxies (\S \ref{subsec:evolution_beta} and Figure \ref{fig:z_beta}). 
These results imply that our LAEs have lower dust contents or younger stellar populations 
at higher $z$ and fainter \muv. 

\item 
The \ew\ values span the range of $\approx 5$ to $240$ \AA\ or larger, 
and the \ew\ distribution can be well fitted by the exponential law, 
$N= N_{\rm0}$ exp($-$\ew/$w_{\rm 0}$) (\S \ref{subsec:measurements_ew} and Figure \ref{fig:ew_comb_distribution}). 
We find that a fainter limiting \muv\ cut increases $w_{\rm 0}$ 
(\S \ref{subsec:cuts_effects} and Figure \ref{fig:scale_lengths_different_cuts}).
These results indicate that selection functions affect $w_{\rm 0}$, 
and care must be taken for the interpretation of the \ew\ distribution, 
its redshift evolution, and their comparisons with previous works.
Taking these effects into account, 
we find that our $w_{\rm 0}$ values are consistent with those in the literature 
within $1\sigma$ uncertainties  at $2.9 < z < 6.6$ at a given \muv\ threshold 
(\S \ref{subsec:evolution_ew} and Figure \ref{fig:evolution_scale_lengths}). 
Given large error bars in our $w_{\rm 0}$ values, 
our data points alone cannot conclude if there exits 
a redshift evolution of $w_{\rm 0}$. 
We need a large sample of LAEs for a definitive conclusion.

\item 
We presented  $12$ LAEs with \ew\ $>200$ \AA\ 
above $1\sigma$ uncertainties (\S \ref{sec:large_ew}, Table \ref{tab:large_ew_LAEs}). 
Among these objects, two LAEs have signatures of merger or AGN activity
indicating that part of the \lya\ emission is contributed from non-star-forming activity. 
For the remaining $10$ LAEs without signatures of mergers or AGNs, 
we constrain stellar ages and metallicities 
based on comparisons between observed \ew\ values 
with stellar synthesis models of \cite{schaerer2003} and \cite{raiter2010} 
under the assumption that all the \lya\ emission originates from star-forming activity. 
We find that these very large \ew\ can be reproduced 
by a recent burst of star formation ($\approx$ 10 Myr) with $Z \lesssim 0.02 \ Z_{\rm \odot}$, 
or by a stellar population younger than $\approx$ 100 Myr (also with $Z \lesssim 0.02 \ Z_{\rm \odot}$) 
for a constant star formation history. 
To put it in another way, 
the very large \ew\ values can be explained 
without invoking PopIII stars or  extremely top-heavy IMFs. 
Alternatively, these very large \ew\ can be also 
explained by, for example, anisotropic radiative transfer effects, 
fluorescence by hidden AGN or QSO activity, 
and/or gravitational cooling. 
 
\end{itemize}

These possible scenarios for very large \ew\ LAEs 
are also invoked to explain \lya\ halo properties presented in L17. 
Thus, in conjunction with our \ew\ and \lya\ halo properties (L17), 
future H$\alpha$ emission line observations 
with, for example, MOSFIRE on Keck and 
{\it The James Webb Space Telescope} (JWST), 
will be very useful to put tighter constraints on these scenarios (L17, \citealt{cantalupo2017, mas-ribas2017}).

\begin{acknowledgements}
This research has been produced within the FOGHAR ANR project 
ANR-13-BS05-110 and the Labex LIO (Lyon Institute of Origins) 
of the Programme Investissements d'Avenir ANR-10-LABX-66.
TH acknowledges the JSPS Research Fellowship for Young Scientists. 
TG is grateful to the LABEX Lyon Institute of Origins (ANR-10-LABX-0066) of the Universit\'e de Lyon for its financial support within the program "d'Avenir" (ANR-11-IDEX-0007) of the French government operated by the National Research Agency (ANR).
JR acknowledges support from the ERC starting grant 336736-CALENDS.
JS thanks the ERC Grant agreement 278594-GasAroundGalaxies. 
RAM acknowledges support by the Swiss National Science Foundation.
JB acknowledges support by Funda{\c c}{\~a}o para a Ci{\^e}ncia e a
Tecnologia (FCT) through national funds (UID/FIS/04434/2013) and Investigador FCT
contract IF/01654/2014/CP1215/CT0003., and by FEDER through COMPETE2020 (POCI-01-0145-FEDER-007672).
SC gratefully acknowledges support from Swiss National Science Foundation grant PP00P2$\_$163824.
We  acknowledge Akio K. Inoue for providing us with the results of his \lya\ transmission shortward of the line. 
We are grateful to Nobunari Kashikawa and Zhenya Zheng for providing us with their data. 
We thank Masami Ouchi, Yuichi Matsuda, Takatoshi Shibuya,  Ken Mawatari, and Kohei Ichikawa for useful discussions. 
\end{acknowledgements}

\bibliographystyle{aa}
\bibliography{hashimoto2017.bib}

\begin{appendix} \section{Summary of the public HST data}

\begin{table*}
\centering
\caption{Summary of the public HST data}
\begin{tabular}{llcccc}
\hline
Instrument/ &  Filter  & Effective$^{a}$ & $2 \sigma^{b}$ & $5 \sigma^{b}$ & Area \\
camera      &           & wavelengths    &  depth               & depth               & \\
                  &             & (\AA)            & (AB mag)           & (AB mag)         & (arcmin$^{2}$) \\
\hline
ACS/WFC3 & F775W & $7693$ & $30.5$ & $29.5$ & $11.4^{e}$ \\
ACS/WFC3 & F850LP & $9055$ & $29.9$ & $28.9$ & $11.4^{e}$ \\
WFC3/IR$^{c}$ & F105W     & $10550$ & $31.1$ & $30.1$ & $4.6$ \\
$^{d}$         &                  &                 & $29.7$ & $28.7$ & $6.8^{f}$ \\
WFC3/IR$^{c}$ & F125W     & $12486$ & $30.7$ & $29.7$ & $4.6$ \\
$^{d}$         &                  &                 & $29.6$ & $28.6$ & $6.8^{f}$ \\
WFC3/IR$^{c}$ & F140W     & $13923$ & $30.8$ & $29.8$ & $4.6$ \\
WFC3/IR$^{c}$ & F160W     & $15370$ & $30.2$ & $29.2$ & $4.6$ \\
$^{d}$         &                  &                 & $29.2$ & $28.2$ & $6.8^{f}$ \\
\hline
\end{tabular}
\tablefoot{
Values taken from \cite{rafelski2015}.\\
$^{a}$ Effective wavelength. \\
$^{b}$ Limiting $2$ and $5\sigma$ magnitudes estimated in an aperture radius of $0''.2$. \\
$^{c}$ Deep and narrow IR data from UDF09 and UDF12 surveys 
(\citealt{oesch2010a, bouwens2011, koekemoer2013, ellis2013}). \\
$^{d}$ Shallow and wide IR data from CANDELS (\citealt{grogin2011, koekemoer2011}). \\
$^{e}$ Source catalog in \cite{rafelski2015} 
is trimmed to central $11.4$ arcmin$^{2}$ from the original coverage of $12.8$ arcmin$^{2}$ 
(see table comments in Table 1 in \citealt{rafelski2015}). \\
$^{f}$ Source catalog in \cite{rafelski2015} is trimmed to $4.6$ arcmin$^{2}$ 
from the original coverage of 6.8 arcmin$^{2}$. 
}
\label{tab:hst_data}
\end{table*}

\end{appendix}

\end{document}